\documentclass[a4paper,11pt]{article}
\usepackage[toc,page]{appendix}
\usepackage{graphicx}
\usepackage{colortbl}
\usepackage{amsmath}
\usepackage{subfigure}
\usepackage{mathtools}

\DeclarePairedDelimiterX\braket[2]{\langle}{\rangle}{#1 \delimsize\vert #2}
\usepackage[utf8]{inputenc}
\usepackage{float}
\usepackage{multirow}
\usepackage[normalem]{ulem}
\usepackage{epstopdf}
\usepackage{amsfonts,amsmath,amssymb}
\usepackage{hyperref}

\usepackage{epsfig,multicol,bbm}
\usepackage{color}
\usepackage[dvipsnames]{xcolor}

\definecolor{darkblue}{rgb}{0.0, 0.0, 0.55}
\definecolor{grey}{rgb}{0.57, 0.64, 0.69}
\definecolor{lightbrown}{rgb}{0.71, 0.4, 0.11}

\newcommand{\be}{\begin{equation}}
\newcommand{\ee}{\end{equation}}

\date{}

\newcommand\fverb{\setbox\pippobox=\hbox\bgroup\verb}
\newcommand\fverbit{\egroup\item[\fbox{\unhbox\pippobox}]}

\newbox\pippobox
\begin{document}
\title{\bf Masses of doubly heavy tetraquarks $QQ\bar{n}\bar{q}$ with $J^{P}=1^{+}$}
\author{Di Gao, Duojie Jia\thanks{Electronic address: jiadj@nwnu.edu.cn},
Yan-Jun Sun
\\
Institute of Theoretical Physics, College of Physics and
Electronic Engineering\\ Northwest Normal University, Lanzhou, 730070,China,\\
}
\maketitle
\begin{abstract}
We apply the method of QCD sum rules to study the doubly heavy tetraquark states $QQ\bar{q}\bar{n}$ with spin-parity $J^{P}=1^{+}$ and strangeness  $S=0, -1$ using careful estimates of the Borel and threshold parameters involved. Masses of the doubly bottom and charmed tetraquarks with isospin $I=0,1/2, 1$ are computed precisely via taking into account multifarious condensates up to dimension $10$. Comparing with the two-heavy meson thresholds, we find that all nonstrange doubly-bottom tetraquarks and a doubly-charmed tetraquarks associted with $J_{3}$ with $J^{P}=1^{+}$ are stable against strong decay into two bottom mesons while a doubly-charmed tetraquarks associated with current $J_{2}$ is unstable against strong decay. By the way, weak decay widths of the doubly bottom tetraquarks are also given.
\end{abstract}
\maketitle
\section{Introduction}
In recent years, a large number of unknown strongly-interacting paricles such as X, Y and Z have been discovered experimentally. Compared with the conventional quark-antiquark mesons and three-quark baryons, these XYZ particles are more difficult to identify due to their potential possibility of mixing exotic multiquark components in them, so understanding these particles via exotic multiquarks has attracted much attention\cite{ExoRev}. In 2020 the LHCb collaboration reported the observation of two exotic structures in the
di-$J/\Psi$ invariant mass spectrum \cite{LHCb2020X69}. One narrow structure of the resonances around $6.9$ GeV, denoted as $X(6900)$, fits to a fully charmed tetraquark $T_{cc\bar{c}\bar{c}}$ and has the measured mass and width
\begin{center}
$M[X(6900)] = 6905 \pm 11 \pm 7 \,\rm{MeV}$,
$\Gamma[X(6900)] = 80 \pm 19 \pm 33 \,\rm{MeV}$.
\end{center}

Very recently, the LHCb collaboration \cite{LHCb2021T} reported important observation of a doubly charmed tetraquark containing two charm quarks, an anti-u and an anti-d quark, using the LHCb-experiment data at CERN, which manifests itself as a narrow peak in the mass spectrum of $D^{0}D^{0}\pi^{+}$ mesons just below the $D^{*+}D^{0}$ mass threshold. This invite quantitative study of mass spectroscopy of the multiquark hadrons and rises issue as if there are (strongly) stable charmed tetraquark $T_{cc}$. For the nonstrange tetraquark $T_{cc}$, most of mass computations \cite{KR:pr2017,Eichten:pr2017,Luo:2017eub,Mehen:D17,M3H:D21,M3H:D22} predict masses around $3.9-4.1$ GeV, above the $D^{*+}D^{0}$ mass threshold $3876$ MeV. On the other hand, given the measured mass $M(\Xi_{cc})=3621.55 \pm 0.53$ MeV of doubly charmed baryon $\Xi_{cc}^{++}(3620)=ccu$ discovered by the LHCb in 2020\cite{LHCb20J}, a simple native sum rule $M(cc\bar{q}\bar{q})$=$2M(ccu)$-$M(cc\bar{c}\bar{c})/2$ predicts the mass of the nonstrange cc tetraquark $T_{cc}$ to be around $3800.1\pm 12.1$ MeV, which is below the $D^{*}D^{0}$ mass threshold. In the past thirty years, doubly heavy tetraquarks have been studied extensively \cite{Ader:D82,Zouzou:C86,Manohar:B93,Bicudo:D15,Bicudo:D17,Mathur:D19S, LuCD:D20}. For recent review, see Refs. \cite{AliBk,Upd2022}.

In this work, we perform a mass analysis of the doubly bottom tetraquarks $bb\bar{n}\bar{n}$, $bb\bar{n}\bar{s}$ and their charm partners $cc\bar{n}\bar{n}$ using the QCD sum rule approach, where the light quarks $n$($=u, d$) can be the up or down quark. A quantitative mass predictions are given for four types of tetraquarks with the mass around $10.3$ GeV for nonstrange states($I = 1$) and $10.5$GeV for strange partners($I = \frac{1}{2}$). The masses of doubly charmed partners $cc\bar{n}\bar{n}$($I =1$) are around $3.8$ GeV, very close to the $D^{*}D^{0}$ mass threshold. In mass analysis, the Borel parameter $M_{b}^{2}$ is confined to the range $[15,20]\,\rm{GeV}^2$ to make sure that the pole contribution dominate at the phenomenological side, and the operator product expansion(OPE) convergents at the quark-gluon side. We also compute weak decay widths of the doubly bottom tetraquarks.

This Letter is organized as follows: after introduction, we outline the QCD sum rule approach for the doubly heavy(DH) tetraquark $QQ\bar{n}\bar{n}$, $QQ\bar{n}\bar{s}$ in Sect. II and in the Sect. III we perform numerical computations of the masses for them in details, with weak decay widths of the doubly bottom tetraquark $bb\bar{q}\bar{n}$ given. The Letter ends with summary in Sect. IV.

\section{QCD sum rule analysis}
In exploring hadron nature at low energy scale, one of successful non-perturbative QCD methods is QCD sum rules\cite{Bicudo:D15,Bicudo:D17}. This method has late been applied to study multifarious hadrons [{\color{blue}17-24}]. In QCD sum rules one uses the quark-hadron duality to balance the (integrated) correlation function.
\begin{equation}
\begin{aligned}
\Pi_{\mu\nu}(q^{2}) &\equiv i \int{ d^{4}x e^{iqx}\langle0| T[J_{\mu}(x)J^{\dagger}_{\nu}(0)] |0\rangle}\\
&=(\frac{q_{\mu}q_{\nu}}{q^{2}}-g_{\mu\nu})\Pi(q^{2}).
\end{aligned}
\end{equation}%
In order to study the DH tetraquarks $ bb\bar{n}\bar{q}(I=0, 1, \frac{1}{2})$,  one constructs the four-quark $QQ\bar{n}\bar{q}$($n$=$u$ and $d$,  $q$=$n$ and $s$) interpolating currents in the  ``diquark-antidiquark'' configuration and considers the Pauli principle to enable all diquark fields to have certain color and spin-flavor structure, composing the tetraquark operator with certain quantum number $J^{P}$. The interpolating currents with $J^{PC}=1^{+}$ for the $bb\bar{q}\,\bar{q}$ tetraquark are[{\color{blue}25}]
\begin{equation}
J_{1}=Q_{a}^{T} C \gamma_{\mu} \gamma_{5} Q_{b}\left(\bar{q}_{a} C \bar{q}_{b}^{T}+\bar{q}_{b} C \bar{q}_{a}^{T}\right)
\end{equation}
\begin{equation}
J_{2}=Q_{a}^{T} C \sigma_{\mu \nu} \gamma_{5} Q_{b}\left(\bar{q}_{a} \gamma^{\nu} C \bar{q}_{b}^{T}-\bar{q}_{b} \gamma^{\nu} C \bar{q}_{a}^{T}\right)
\end{equation}
\begin{equation}
J_{3}=Q_{a}^{T} C \gamma_{\mu} Q_{b}(\bar{q}_{a} \gamma_{5} C \bar{q}_{b}^{T}-\bar{q}_{b} \gamma_{5} C \bar{q}_{a}^{T})
\end{equation}
\begin{equation}
J_{4}=Q_{a}^{T} C \sigma_{\mu \nu} Q_{b}(\bar{q}_{a} \gamma^{\nu} \gamma_{5} C \bar{q}_{b}^{T}-\bar{q}_{b} \gamma^{\nu} \gamma_{5} C \bar{q}_{a}^{T}).
\end{equation}
Here, the current $J_{1}$ in Eq. (2) and $J_{2}$ in Eq. (3) belong to symmetric flavor structure and form the $I=1$ isotriplet($\bar{u}\bar{u}$, $\bar{d}\bar{d}$, $(\bar{u}\bar{d}+\bar{d}\bar{u})/\sqrt{2}$) while the current $J_{3}$ in Eq.(4) and $J_{4}$ in Eq. (5) belong to antisymmetric flavor structure and form the $I=0$ isosinglet($(\bar{u}\bar{d}-\bar{d}\bar{u})/\sqrt{2}$). Due to involved four-body QCD interaction, spin and color configurations of DH tetraquark system via its subsystem $QQ$ and $\bar{q}\bar{q}$ is involved(Appendix A).

At the hadron level, we can express $ \Pi(q^{2}) $ in the form of the dispersion relation with a spectral function $ \rho_{phen}(s) $ :
\begin{equation}
\Pi(q^{2}) = \int_{(2m_{Q}+2m_{q})^{2}}^{\infty}ds\frac{\rho_{phen}(s)}{s-q^{2}-i\varepsilon}
\end{equation}%
with the integration starting from the physical threshold, $(2m_{Q}+2m_{q})^{2}$. Here, the spectral density $\rho_{phen}(s) $ is the imaginary part of the correlating function, $\rho_{phen}(s) = Im\Pi(s)/\pi $ .

A parameterization of one-pole dominance for the lowest state and a continuum contribution for the excited states re-expresses the spectral density in the following form
\begin{equation}
\begin{aligned}
\rho_{phen}(s) &\equiv \frac{1}{\pi}Im\Pi(s) \\
&= \Sigma \delta(s-M_{n}^{2}) \langle0| J |n\rangle \langle n| J^{\dagger} |0\rangle\\
&= f_{x}^{2}\delta (s-M_{x}^{2}) + higher\ states,\\
\end{aligned}
\end{equation}%
where $f_{x}$ is the coupling strength of the hadron with $ J(x) $ in the hadron spectrum expansion, $M_{x}$ is the ground-state mass of hadron and $ J(x) $ contains the contributions of higher states and continuum.
\par
At the quark-gluonic level, Eq. (1) are calculated with the OPE. Performing the Borel transformation both at hadron and quark-gluon levels, one finds
\begin{equation}
\Pi(M_{b}^{2},\infty) \equiv \widehat{B}_{M_{b}^{2}}\Pi(q^{2}) = \int_{(2m_{Q}+2m_{q})^{2}}^{\infty}ds e^{-s/M_{x}^{2}}\rho(s).
\end{equation}%
Approximating the contribution from the continuum states by the spectral density above a threshold value $ s_{0} $, one obtains the sum rule relation
\begin{equation}
\Pi(M_{b}^{2},s_{0}) \equiv f_{x}^{2}e^{-M_{x}^{2}/M_{b}^{2}} = \int_{(2m_{Q}+2m_{q})^{2}}^{s_{0}}dse^{-s/M_{b}^{2}}\rho(s),
\end{equation}%
from which one can extract the hadron mass $M_{x}$ of the lowest-lying resonance to be
\begin{equation}
M_{x}^{2}(M_{b}^{2},s_{0}) = \frac{\frac{\partial}{\partial(-\frac{1}{M_{b}^{2}})}{\Pi(M_{b}^{2},s_{0})}}{\Pi(M_{b}^{2},s_{0})} = \frac{\int_{(2m_{Q}+2m_{q})^{2}}^{s_{0}}dse^{-s/M_{b}^{2}}\rho(s)s}{\int_{(2m_{Q}+2m_{q})^{2}}^{s_{0}}dse^{-s/M_{b}^{2}}\rho(s)}.
\end{equation}%

To find heavy tetraquark mass, we have to compute the integration in RHS of Eq. (8). For this, we consider all Feynman diagrams of the quark, gluon and mixed condensates up to dimension 10, and plot all Feynman diagrams for the two-point functions of the tetraquark currents in FIG.1. In the case of the tetraquark $QQ\bar{n}\bar{n}$($J_{1\mu}^{+}$) with $\bar{n}=\bar{u},\bar{d}$, as an example, we derive the explicit form of spectral densities, as shown in Appendix B.

\begin{figure}[H]\hspace{0.4cm}
\centering
\subfigure[]{\includegraphics[width=1\columnwidth]{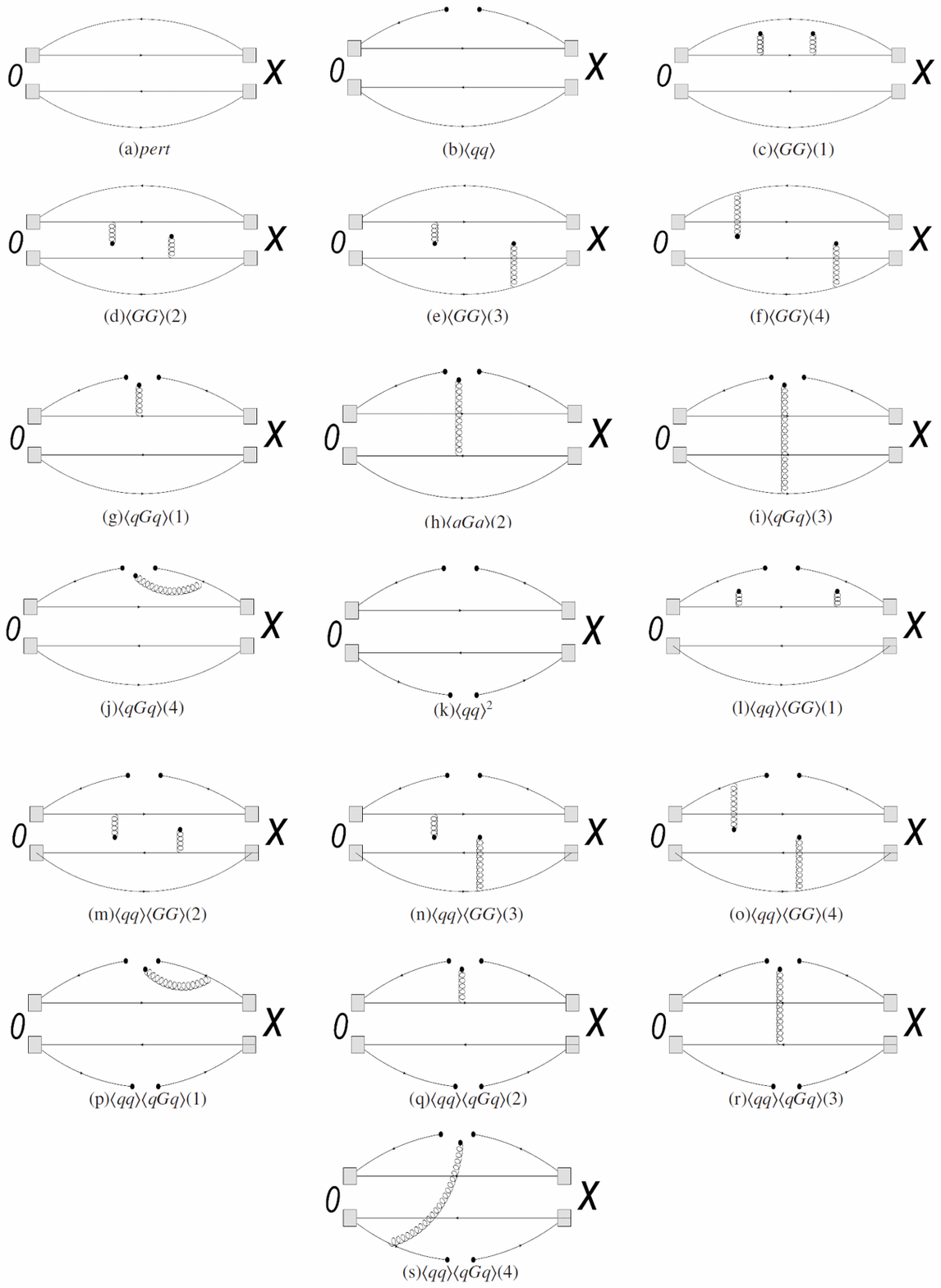}}
\caption{Feynman diagrams for the two-point function of the tetraquark current $\Pi_{\mu\nu}$. Squares represent events, straight line represent heavy quarks, curves represent light quarks, and  helical lines represent gluons. The letters a, b, \ldots, s are used to count the number of condensates, while the numbers 1,2,3,4 and 5 represent different condensates of the same form.}
\end{figure}
\par
Next, we give a detailed analysis using $bb\bar{n}\bar{q}$ as an example, i.e., $Q=b$, followed by the results for $cc\bar{n}\bar{q}$ ($Q=c$).

\section{Numerical analysis and discussions}
Before numerical computation we use the following inputs of parameters for quark masses and various QCD condensates[{\color{blue}26-32}]:
\begin{center}
$m_{b} = 4.18_{-0.03}^{+0.04} \text{ GeV} $,
$m_{c} = 1.27 \pm 0.02 \text{ GeV}$,
$m_{s} = 96_{-4}^{+8} \text{ MeV}$,

$\langle\bar{q}q\rangle = -(240 \pm 10 \text{  MeV})^{3}$,
$\langle\bar{q}g_{s}\sigma \cdot G q\rangle = -M_{0}^{2}\langle\bar{q}q\rangle$,

$\langle g_{s}^{2}GG\rangle=(0.48\pm0.14) \text{ GeV}^{4}$,
$M_{0}^{2} = (0.8 \pm 0.2) \text{ GeV}^{2}$.
\end{center}
which are fixed in whole work. The Borel parameter $M_{b}$ and threshold $s_{0}$ can vary within the appropriate regions, which have to satisfy the standard restrictions from the sum rules computations. The window for $M_{b}^{2} \in [{M_{b}^{2}}_{min},{M_{b}^{2}}_{max}]$ is fixed from the constraints imposed on the pole contribution (PC) which determines ${M_{b}^{2}}_{max}$ and the convergence ratio $R({M_{b}^{2}}_{min})$ necessary to find ${M_{b}^{2}}_{min}$. The definition for the PC is
\begin{equation}
PC \equiv \frac{\Pi({M_{b_{max}}^{2}},s_{0})}{\Pi({M_{b_{max}}^{2}},\infty)} = \frac{\int_{(2m_{b}+2m_{q})^{2}}^{s_{0}}dse^{-s/M_{b}^{2}}\rho(s)}{\int_{(2m_{b}+2m_{q})^{2}}^{\infty}dse^{-s/M_{b}^{2}}\rho(s)},
\end{equation}
and that for $R({M_{b}^{2}}_{min})$ is
\begin{equation}
R({M_{b}^{2}}_{min}) \equiv \frac{\Pi^{DimN}({M_{b}^{2}}_{min},s_{0})}{\Pi({M_{b}^{2}}_{min},s_{0})},
\end{equation}
where $\Pi^{DimN}({M_{b}^{2}}_{min},s_{0})$ is the contribution of the higher orders.

We take into account all of the aforementioned constraints to carry out the numerical analysis, and determine the optimal regions for $M_{b}^{2}$ and $s_{0}$. During the search for the Borel parameter $M_{b}^{2}$ and the continuum threshold parameter $s_{0}$ the following criteria are used:

(1) Pole dominates at the phenomenological(hadron) side.

(2) The OPE is convergent.
\par
(3) Borel platforms emerge.

\begin{figure}[H]\hspace{0.4cm}
\centering
\subfigure{\includegraphics[width=0.8\columnwidth]{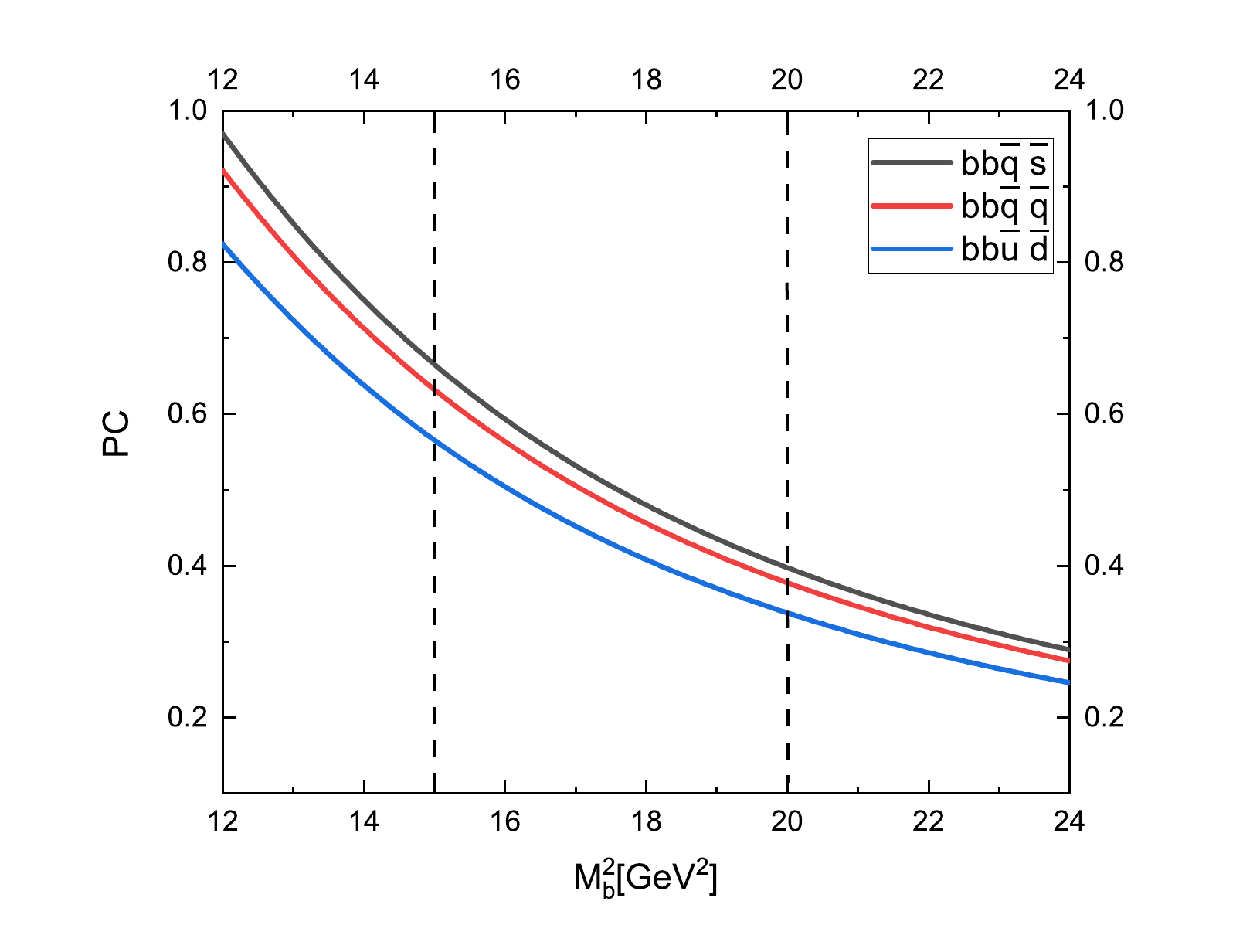}}
\caption{The PC($J_{1}$ for $bb\bar{n}\bar{n}$ and $bb\bar{n}\bar{s}$, $J_{3}$ for $bb\bar{u}\bar{d}$), defined in Eq.(11), as a function of the Borel parameter $M_{b}$. The curve is obtained by taking $s_{0}(bb\bar{u}\bar{d})$=$133.8\,\rm{GeV}^{2}$ , $s_{0}(bb\bar{n}\bar{n})$=$134.3\,\rm{GeV}^{2}$ and $s_{0}(bb\bar{n}\bar{s})$=$138.5\,\rm{GeV}^{2}$.}
\end{figure}

For the infinity ($``\infty"$) of the denominator of PC in Eq. (11), one has to regularize the integration over all excited states of $bb\bar{q}\bar{q}$. Phyisically, it is enough to find an appropriate upper limit of the integral to replace the infinity. Then, one can find this upper limit with the help of a set of mass inequalities, $max\{m_{bb\bar{q}{\bar{q}}}\}$ $<$ $min_{Q\bar{Q}}\{min\{m_{bb\bar{q}{\bar{q}}Q\bar{Q}}\}\}$ $<$ $max_{Q\bar{ Q}}\{min\{m_{bb\bar{q}{\bar{q}}Q\bar{Q}}\}\}$ $\leq$  $min\{m_{bb\bar{q}{\bar{q}}b\bar{b}}\}$ $<$
$2m_{\bar{B}/\bar{B}_{s}}+m_{\eta_{b}/h_{b}/\Upsilon/\chi_{b}}$ $\approx$ $20\ \rm{GeV}$ (Appendix C), which rises from the features of the QCD quantum vacuum (containing sea-quarks) and color confining of QCD. Finally, we can estimate the lower limits of the PC for every $M_{b}^{2}$ in FIG. 2.

\begin{figure}[H]\hspace{0.4cm}
\centering
\subfigure{\includegraphics[width=0.8\columnwidth]{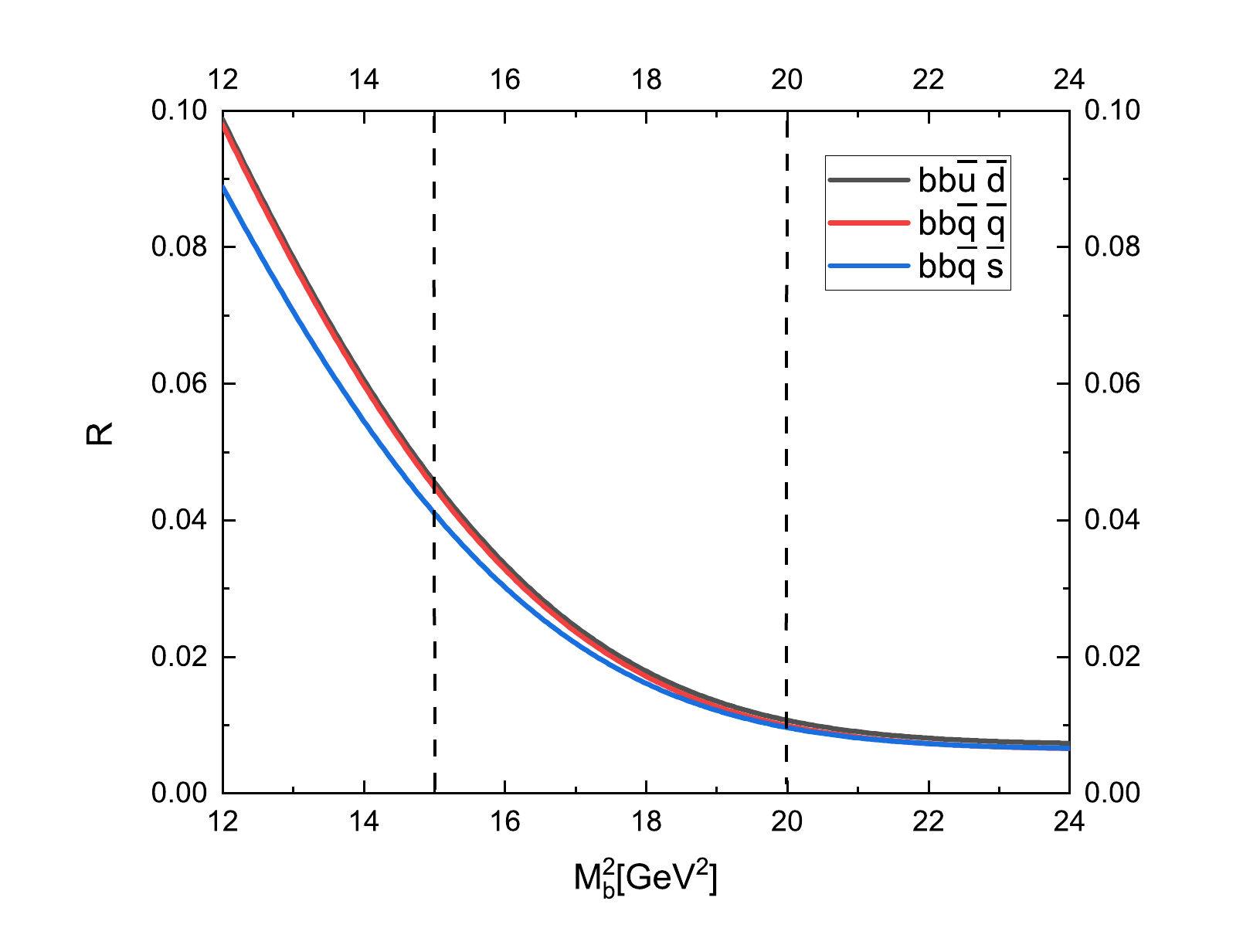}}
\caption{The convergence ratio R($J_{1}$ for $bb\bar{n}\bar{n}$ and $bb\bar{n}\bar{s}$, $J_{3}$ for $bb\bar{u}\bar{d}$), defined in Eq.(12), as a function of the Borel parameter $M^{2}_{b}$. The curve is obtained by taking $s_{0}(bb\bar{u}\bar{d})$=$133.8\,\rm{GeV}^{2}$,$s_{0}(bb\bar{n}\bar{n})$=$134.3\,\rm{GeV}^{2}$ and $s_{0}(bb\bar{n}\bar{s})$=$138.5\,\rm{GeV}^{2}$.}
\end{figure}

We compute the PC and find it to be in the ranges
$ 39.76 $\%$ < PC(bb\bar{n}\bar{s}(J_{1})) < 66.57$\%$ $
($ 37.77$\%$ < PC(bb\bar{n}\bar{n}(J_{1})) < 63.24$\%$ $)
($ 33.80$\%$ < PC(bb\bar{u}\bar{d}(J_{3})) < 56.58$\%$ $)
in the regions  $15\,\rm{GeV}^{2}$ $<$ $M_{b}^{2}$ $<$ $20\,\rm{GeV}^{2}$ with $J^{P}=1^{+}$, as shown in FIG. $2$. We also calculate the ratio R and find it to be in the ranges
$ 0.96$\%$ < R(bb\bar{n}\bar{s}(J_{1})) < 4.11$\%$ $
($ 0.99$\%$ <R(bb\bar{n}\bar{n}(J_{1})) < 4.48$\%$ $)
($ 1.07$\%$ <R(bb\bar{u}\bar{d}(J_{3})) < 4.56$\%$ $)
in the regions  $15\,\rm{GeV}^{2}$ $<$ $M_{b}^{2}$ $<$ $20\,\rm{GeV}^{2}$(shown in FIG. $3$). Similar calculations yield the following ranges of PC and R for other configurations,

\begin{equation}
\centering
\begin{aligned}
bb\bar{n}\bar{n}(J_{2})\quad:\quad 36.07\% < PC < 65.21\% , 0.94\% < R < 4.14\% ;\\
bb\bar{n}\bar{s}(J_{2})\quad:\quad 37.90\% < PC < 65.32\% , 0.93\% < R < 4.09\% ;\\
bb\bar{n}\bar{s}(J_{3})\quad:\quad 35.72\% < PC < 59.57\% , 0.97\% < R < 4.32\% ;\\
bb\bar{u}\bar{d}(J_{4})\quad:\quad 32.02\% < PC < 55.69\% , 1.09\% < R < 5.00\% ;\\
bb\bar{n}\bar{s}(J_{4})\quad:\quad 38.01\% < PC < 65.62\% , 0.93\% < R < 4.12\% .\\
\end{aligned}
\centering
\end{equation}

Putting all together, one sees that the listed ranges turn out to be appropriate in view of Borel platforms. The optimal ranges we then obtain are:
\begin{equation}
\begin{aligned}
bb\bar{u}\bar{d}\quad:\quad M_{b}^{2} = (15-20)\,\rm{GeV}^{2}, s_{0} = (131.3-136.3)\,\rm{GeV}^{2},\\
bb\bar{n}\bar{n}\quad:\quad M_{b}^{2} = (15-20)\,\rm{GeV}^{2}, s_{0} = (131.8-136.8)\,\rm{GeV}^{2},\\
bb\bar{n}\bar{s}\quad:\quad M_{b}^{2} = (15-20)\,\rm{GeV}^{2}, s_{0} = (136.0-141.0)\,\rm{GeV}^{2}.
\end{aligned}
\end{equation}
To reduce the uncertainty from the PC and R, we plot the mass dependence of the tetraquarks $bb\bar{n}\bar{n}$ and $bb\bar{n}\bar{s}$ upon $M_{b}^{2}$ and $s_{0}$ in FIG. 4 and FIG. 5.

\begin{figure}[H]\hspace{0.4cm}
\centering
\subfigure{\includegraphics[width=0.6\columnwidth]{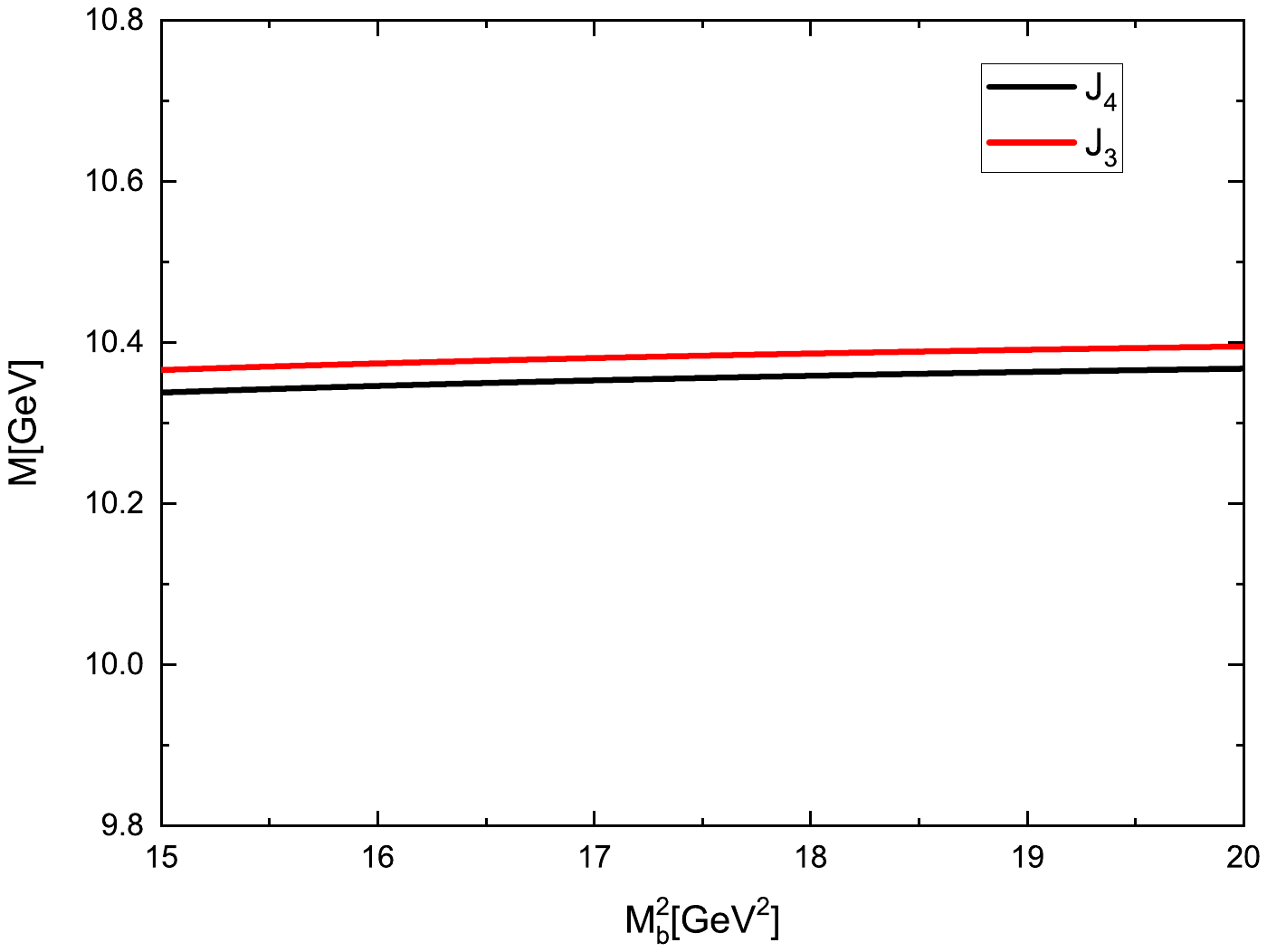}}
\subfigure{\includegraphics[width=0.6\columnwidth]{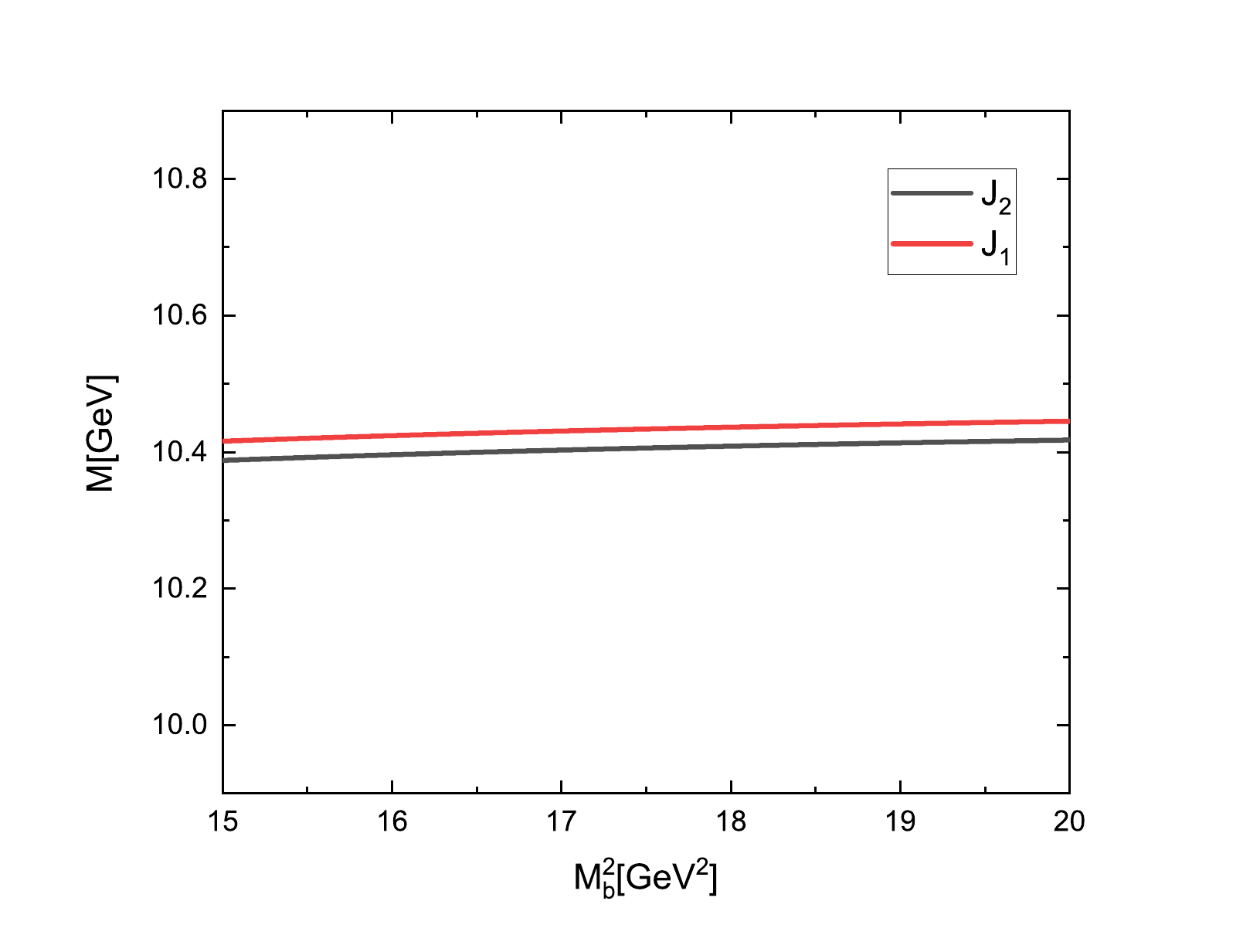}}
\subfigure{\includegraphics[width=0.6\columnwidth]{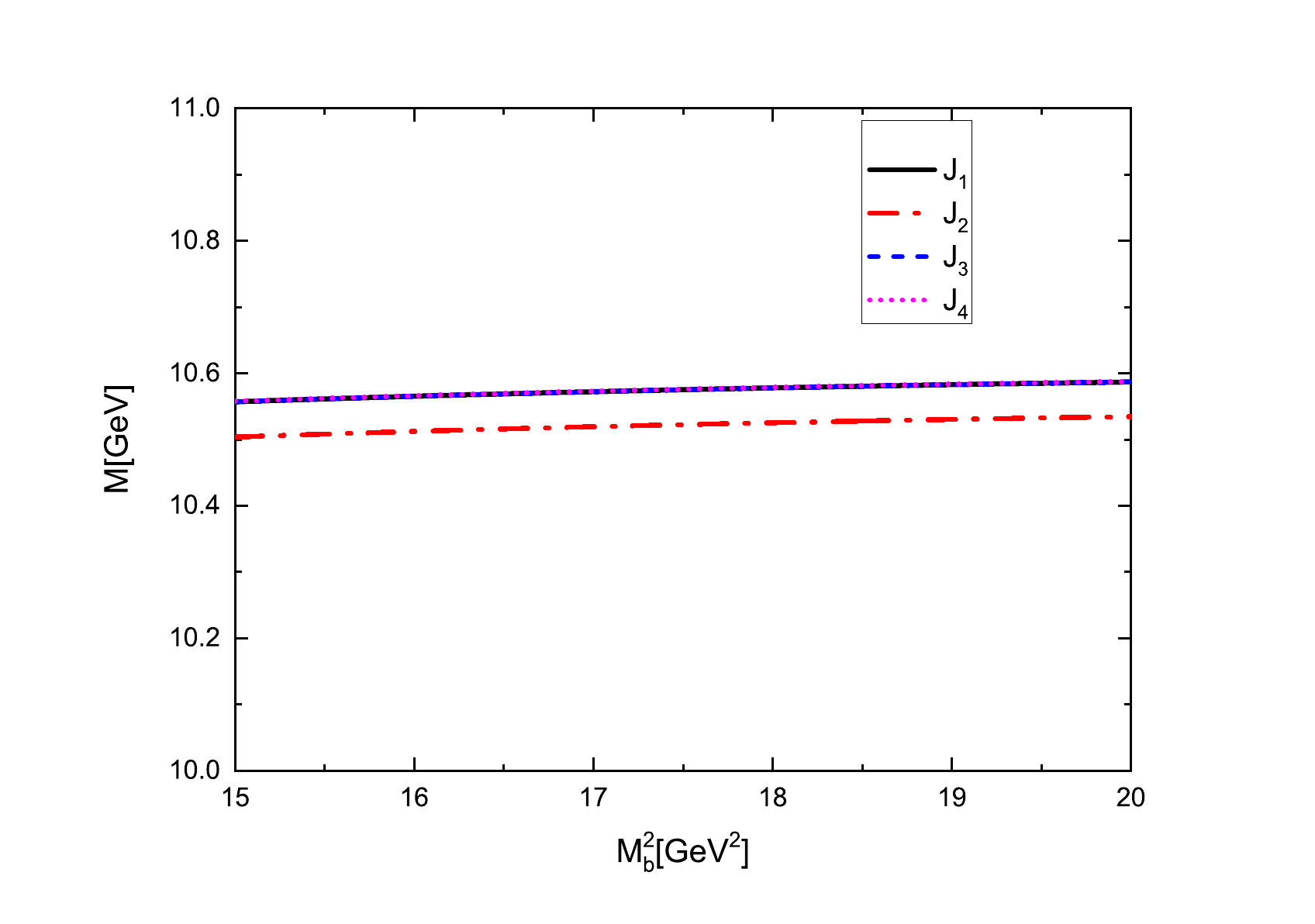}}
\caption{Dependence of the mass of the tetraquarks $bb\bar{u}\bar{d}$ (a), $bb\bar{n}\bar{n}$ (b) and $bb\bar{n}\bar{s}$(c) upon the Borel parameters $M_{b}^{2}$. The curves are obtained with $s_{0}(bb\bar{u}\bar{d})$=$133.8.0\,\rm{GeV}^{2}$, $s_{0}(bb\bar{n}\bar{n})$=$134.3.0\,\rm{GeV}^{2}$ and $s_{0}(bb\bar{n}\bar{s})$=$138.5\,\rm{GeV}^{2}$.}
\end{figure}

\begin{figure}[H]\hspace{0.4cm}
\centering
\subfigure{\includegraphics[width=0.6\columnwidth]{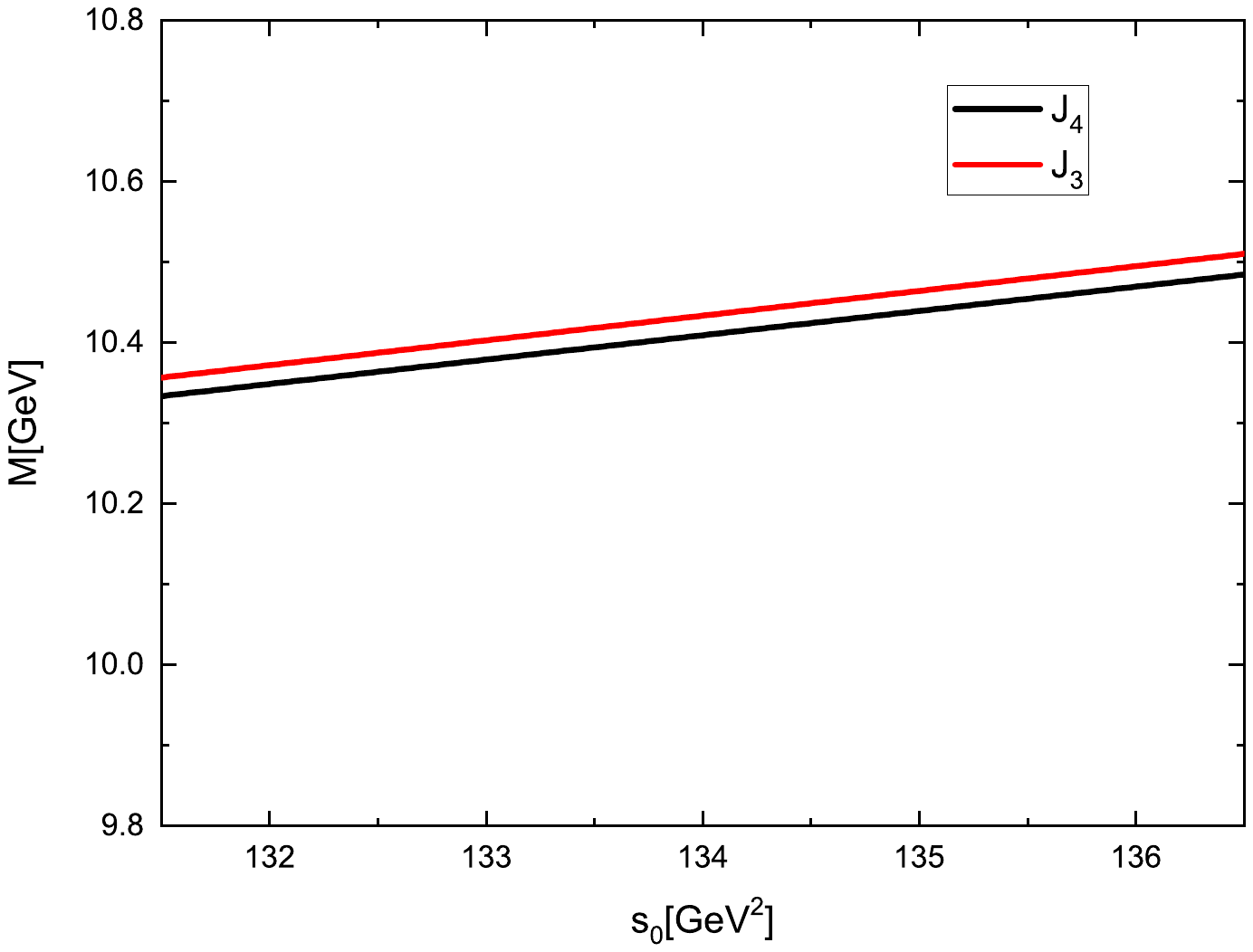}}
\subfigure{\includegraphics[width=0.6\columnwidth]{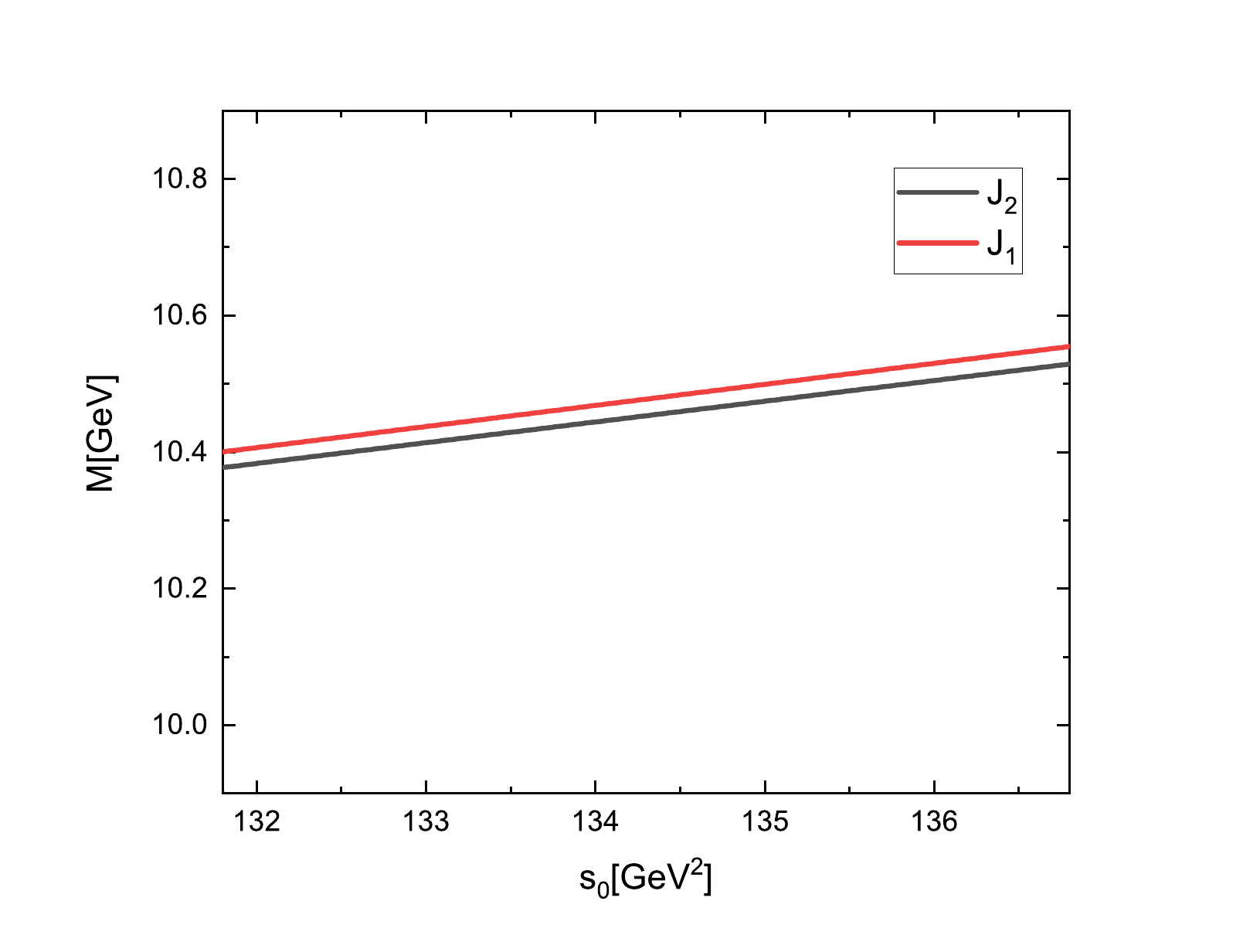}}
\subfigure{\includegraphics[width=0.6\columnwidth]{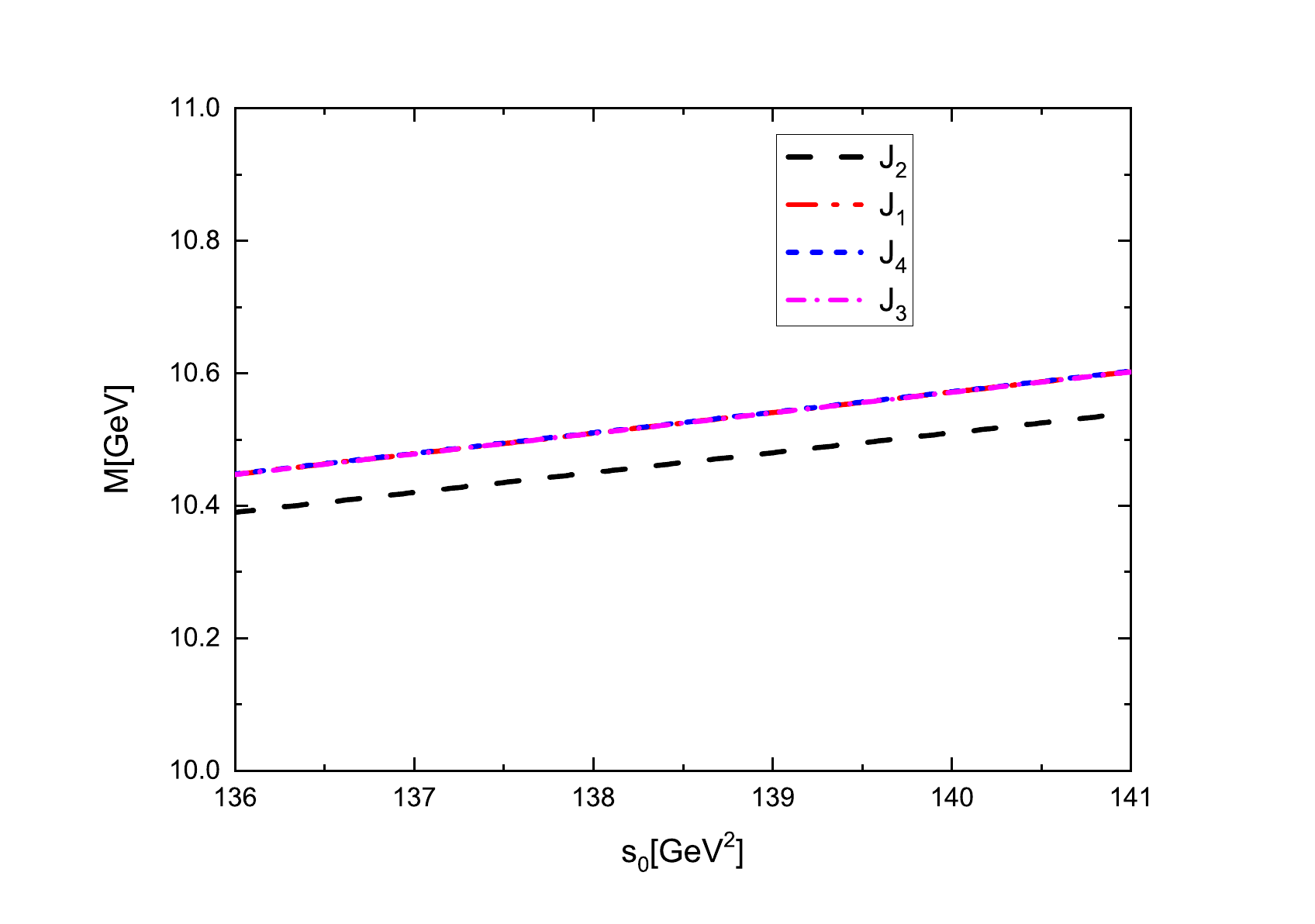}}
\caption{Dependence of the mass of the tetraquarks $bb\bar{u}\bar{d}$ (a), $bb\bar{n}\bar{n}$ (b) and $bb\bar{n}\bar{s}$(c) upon the threshold parameters $s_{0}$. The curves are obtained with $M_{b}^{2}=17\,\rm{GeV}^{2}$.}
\end{figure}

In FIG. 4, we plot the mass prediction of $bb\bar{n}\bar{q}$ depending upon the Borel parameter $M_{b}^{2}$, which confirms the values used in Eq. (14). It is seen that the $M_{b}^{2}-$dependence of the mass is very weak: the computed masses of $bb\bar{n}\bar{s}$ show a high stability against varying of $M_{b}^{2}$ in the optimized working interval. In FIG. 5, we plot the mass prediction of the strange state $bb\bar{n}\bar{q}$ depending upon $s_{0}$. While the computed masses do depend on the continuum threshold $s_{0}$, which yields a main part of uncertainties( due to uncertainty of $s_{0}$ ), one can regard, in the light of standard limits acceptable for our computations, that they remain a constant approximately for the chosen intervals of $s_{0}$ in Eq. (12).

\makeatletter\def\@captype{table}\makeatother

\begin{center}
Table \uppercase\expandafter{\romannumeral1}: Computed masses (in GeV) of the nonstrange doubly-bottom tetraquarks with $J^{P}=1^{+}$, including the binding energies relative to two heavy-meson decays and computed decay thresholds in this work.
\begin{tabular}{p{2.5cm}<{\centering}|p{2cm}<{\centering}p{2cm}<{\centering}|p{2cm}<{\centering}p{2cm}}
\hline
State&\multicolumn{2}{c}{$bb[\bar{u}\bar{d}]$}&\multicolumn{2}{c}{$bb\{\bar{n}\bar{n}\}$}\\
~&$J_{3}$&$J_{4}$&$J_{1}$&$J_{2}$\\
\hline
Our work &$10.380^{+0.03}_{-0.03}$ & $10.353^{+0.04}_{-0.03}$ & $-$ & $10.403^{+0.03}_{-0.04}$\\
Decay (GeV) &\multicolumn{2}{c}{$B^{-}\bar{B}^{*0}$ $(10.520\pm 0.032)$}&\multicolumn{2}{c}{$B^{-}\bar{B}^{*0}$($10.520\pm 0.032)$} \\
$E_{binding}$ (MeV) &$-140\pm 62 $ & $-167_{-62}^{+72}$ &$-$ &$-117_{-72}^{+62} $\\
$[{\color{blue}33}]$ &$10.2\pm0.3$ &$10.2\pm0.3$ &$10.2\pm0.3$ & $-$  \\
$[{\color{blue}6}]$ &\multicolumn{2}{c}{10.482} & \multicolumn{2}{c}{10.681} \\
$[{\color{blue}34}]$&\multicolumn{2}{c}{$10.471\pm0.025$}&\multicolumn{2}{c}{$10.671\pm0.025$}\\
$[{\color{blue}5}]$&10.821&10.686&\multicolumn{2}{c}{10.875}\\
$[{\color{blue}35}]$&\multicolumn{2}{c}{10.690}&\multicolumn{2}{c}{$-$}\\
$[{\color{blue}36}]$&\multicolumn{2}{c}{10.586}&\multicolumn{2}{c}{$-$}\\
$[{\color{blue}37}]$&\multicolumn{2}{c}{10.36}&\multicolumn{2}{c}{$-$}\\
$[{\color{blue}38}]$&10.550&10.951&\multicolumn{2}{c}{10.779}\\
\hline
\end{tabular}
\end{center}

\bigskip
Given all above considerations, we are in the position to compute the masses of the doubly bottom tetraquarks with isospin=1,0 and 1/2 and   $J^{P}=1^{+}$, with strangeness=0 and 1. The results obtained are listed collectively in Table \uppercase\expandafter{\romannumeral1} and Table \uppercase\expandafter{\romannumeral2} and compared to other works cited. The binding energy for the decay is obtained by $E_{binding}=m(QQ\bar{n}\bar{q})-[m(Q\bar{n})+m(Q\bar{q})]$.

In Table \uppercase\expandafter{\romannumeral1} and Table \uppercase\expandafter{\romannumeral2}, the central values correspond to $M_{b}^{2}=17\,\rm{GeV}^{2}$, $s_{0}(bb\bar{u}\bar{d})=133.8\,\rm{GeV}^{2}$, $s_{0}(bb\bar{n}\bar{n})=134.3\,\rm{GeV}^{2}$ and $s_{0}(bb\bar{n}\bar{s})=138.5\,\rm{GeV}^{2}$, and the first and second uncertainties are due to the Borel parameter $M_{b}^{2}$ and the threshold parameter $s_{0}$, respectively. In our computations, we have not considered the uncertainty due to other parameters such as $m_{b}$, $m_{q}$, multifarious condensates and so on.

\bigskip

\begin{center}
Table \uppercase\expandafter{\romannumeral2}: Computed masses(in GeV) of the strange doubly-bottom tetraquarks with $J^{P}=1^{+}$, including binding energies relative to two heavy-mesons and computed decay thresholds in this work.
\begin{tabular}{p{2.5cm}<{\centering}|p{2cm}<{\centering}p{2cm}<{\centering}|p{2cm}<{\centering}p{2cm}}

\hline
State&\multicolumn{2}{c}{$bb[\bar{n}\bar{s}]$}&\multicolumn{2}{c}{$bb\{\bar{n}\bar{s}\}$}\\
~&$J_{3}$&$J_{4}$&$J_{1}$&$J_{2}$\\
\hline
Our work &$10.571^{+0.02}_{-0.03}$&$10.520^{+0.03}_{-0.02}$&$10.570^{+0.03}_{-0.02}$&$10.574^{+0.03}_{-0.02}$\\
Decay(GeV) &\multicolumn{2}{c}{ $\bar{B}B^{*}_{s}(10.590\pm 0.032)$ } & \multicolumn{2}{c}{ $\bar{B}B^{*}_{s}(10.590\pm 0.032)$  } \\
$E_{binding}$(MeV) & $-19^{+52}_{-62}$ & $-70^{+62}_{-52}$  &$-20^{+62}_{-52}$ & $-16^{+62}_{-52}$ \\
$[{\color{blue}33}]$&$10.7\pm0.3$&$10.4\pm0.3$&$10.3\pm0.3$&$10.4\pm0.3$\\
$[{\color{blue}6}]$&\multicolumn{4}{c}{10.643}\\
$[{\color{blue}34}]$&\multicolumn{4}{c}{$10.644\pm0.026$}\\
$[{\color{blue}35}]$&\multicolumn{4}{c}{10820}\\
$[{\color{blue}36}]$&\multicolumn{4}{c}{10.629}\\
$[{\color{blue}37}]$&\multicolumn{4}{c}{10.51}\\
$[{\color{blue}38}]$&10.734&10.897&\multicolumn{2}{c}{11.046}\\
\hline
\end{tabular}
\end{center}

For the systems of the pseudoscalar ($P$) mesons and vector ($V$) mesons ($\bar{%
q}Q$), one can construct the correlation functions,
\begin{eqnarray}
\Pi _{5}^{P}(q) &=&i\int d^{4}xe^{iq\cdot x}\langle 0|T[J_{5}(x)J_{5}^{\dag
}(0)]|0\rangle ,  \label{1} \\
\Pi _{\mu \nu }^{V}(q) &=&i\int d^{4}xe^{iq\cdot x}\langle 0|T[J_{\mu
}(x)J_{\nu }^{\dag }(0)|0\rangle ,  \label{2}
\end{eqnarray}%
with $J_{5}(x)=\bar{q}i\gamma _{5}Q$ and $J_{\mu }(x)=\bar{q}i\gamma _{\mu }Q
$ the respective currents of the heavy mesons $\bar{q}Q$. Then, one can perform OPE upon these two functions up to the mass dimension of eight for the condensation to obtain the Borel transformed correlation functions for both currents $J_{5,\mu }$, as done by Ref. \cite{KL}, for instance. Thus, one can apply the same method of the QCD sum rule to compute the
masses of the heavy mesons. The results are
\begin{eqnarray*}
m(B) &=&5.24\pm 0.012\text{ GeV, }m(B^{\ast })=5.28\pm 0.02\text{ GeV,} \\
m(B_{s}) &=&\text{ }5.31\pm 0.019\text{ GeV, }m(B_{s}^{\ast })=5.35\pm 0.02\
\text{GeV,} \\
m(D) &=&1.87\pm 0.006\text{ GeV, }m(D^{\ast })=1.99\pm 0.002\text{ GeV,}
\end{eqnarray*}%
which yield the (two-meson) mass threshold of the tetraquarks
\begin{eqnarray}
m(B\bar{B}^{\ast }) &=&10.52\pm 0.032\text{ GeV, }m(D\bar{D}^{\ast
})=3.860\pm 0.008\text{ GeV,}  \notag \\
m(\bar{B}B_{s}^{\ast }) &=&10.59\pm 0.032\text{ GeV, }m(\bar{B}_{s}B^{\ast
})=10.59\pm 0.039\text{ GeV.}  \label{threshold}
\end{eqnarray}

In Ref. [{\color{blue}4}], it is suggested that the tetraquark $bb\bar{n}\bar{q}$ decays weakly since it is deeply bounded. Assuming a final state $\bar{B}D$ for weak decay of a given tetraquark $bb\bar{n}\bar{q}$ with a charged weak current giving rise to $e\bar{\nu}_{e}$, $\mu\bar{\nu}_{\mu}$, $\tau\bar{\nu}_{\tau}$, one can use color factor(=3) of $\bar{q}q$ and $\bar{c}s$, a CKM matrix element $|V_{cb}|=0.04$[{\color{blue}39}] and a factor(=2) counting each decaying of $b$ quark to compute its decay rate. The widths for all tetraquark $QQ\bar{n}\bar{q}$ states with $J^{P} = 1^{+}$ are[{\color{blue}4}]
\begin{equation}
\Gamma(bb\bar{n}\bar{q})=\frac{18G_{F}^{2}M(bb\bar{n}\bar{q})^{5}}{192\pi^{3}}F(x)|V_{cb}|^{2},
\end{equation}
in which the kinematic suppression factor $F(x)$ is given by
\begin{equation}
F(x)=1-8x+8x^{3}-x^{4}+12x^{2}\text{ln}\left(\frac{1}{x}\right), x\equiv \left(\frac{M(\bar{B})+M(D)}{M(bb\bar{n}\bar{q})} \right)^{2},
\end{equation}
with $M(\bar{B})$, $M(D)$ and $M(bb\bar{n}\bar{n})$ the masses of the heavy meson $B$, $D$ and the $bb\bar{q}\bar{q}$, respectively. The results obtained thereby are collected in Table \uppercase\expandafter{\romannumeral3}. In obtaining Table \uppercase\expandafter{\romannumeral3}, we have used the following masses $M(bb\bar{n}\bar{q})$ of the initial decaying tetraquark states: ${m_{bb\bar{u}\bar{d}}}_{J_{3}}=10.380\,\rm{GeV}$, ${m_{bb\bar{u}\bar{d}}}_{J_{4}}=10.353\,\rm{GeV}$, ${m_{bb\bar{n}\bar{n}}}_{J_{1}}=10.431\,\rm{GeV}$, ${m_{bb\bar{n}\bar{n}}}_{J_{2}}=10.403\,\rm{GeV}$, ${m_{bb\bar{n}\bar{s}}}_{J_{1}}=10.571\,\rm{GeV}$, ${m_{bb\bar{n}\bar{s}}}_{J_{2}}=10.520\,\rm{GeV}$, ${m_{bb\bar{n}\bar{s}}}_{J_{3}}=10.570\,\rm{GeV}$ and ${m_{bb\bar{n}\bar{s}}}_{J^{-}_{4}}=10.574\,\rm{GeV}$.

\bigskip

\begin{center}
Table \uppercase\expandafter{\romannumeral3}: The decay widths of the tetraquarks $bb\bar{u}\bar{d}, bb\bar{n}\bar{n}$ and $bb\bar{n}\bar{s}$ to $\bar{B}D~$  or $~\bar{B}D_{s}$.

\begin{tabular}{p{2.5cm}<{\centering}p{2.5cm}<{\centering}p{2.5cm}<{\centering}p{2.5cm}<{\centering}}

\hline
Decay channel&Current&Our work ($\rm{GeV}$)& Ref.[{\color{blue}4}]( $\rm{GeV}$)\\
\hline
\multirow{2}{*}{$bb\bar{u}\bar{d}\rightarrow\bar{B}$ $D$}&$J_{3}$ & $17.51\times10^{-13}$ &\multirow{2}{*}{$17.9\times10^{-13}$}\\
&$J_{4}$&$16.85\times10^{-13}$\\
\hline
\multirow{2}{*}{$bb\bar{n}\bar{n}\rightarrow\bar{B}$ $D$}&$J_{1}$&$-$\\
&$J_{2}$&$18.09\times10^{-13}$\\
\hline
\multirow{4}{*}{$bb\bar{n}\bar{s}\rightarrow\bar{B}$ $D_{s}$}&$J_{1}$&$19.46\times10^{-13}$&\multirow{4}{*}{-}\\
&$J_{2}$&$18.00\times10^{-13}$\\
&$J_{3}$&$20.19\times10^{-13}$\\
&$J_{4}$&$19.27\times10^{-13}$\\
\hline
\end{tabular}
\end{center}

Similar analysis applies to the strange partners $bb\bar{n}\bar{s}$ of the above tetraquarks, and one can then compute their decay widths in the channels $bb\bar{u}\bar{d}$/$bb\bar{n}\bar{n}$/$bb\bar{n}\bar{s}$$\rightarrow$$\bar{B}D$$/$$\bar{B}D_{s}$ for the configurations with $J^{P} = 1^{+}$. The computed results are listed collectively in Table \uppercase\expandafter{\romannumeral2}, where all widths are of order of $10^{-13}$ GeV. In both of Tables I and II, the calculated results in Ref. [{\color{blue}4}] are also shown for comparison.
\bigskip

\begin{center}
Table \uppercase\expandafter{\romannumeral4}: Computed masses(in GeV) of the nonstrange doubly charmed tetraquarks with $J^{P}=1^{+}$,and binding energies relative to two-meson decay and computed decay shresholds in this work.
\begin{tabular}{p{2.5cm}<{\centering}|p{2cm}<{\centering}p{2cm}<{\centering}|p{2cm}<{\centering}p{2cm}<{\centering}}
\hline
State &\multicolumn{2}{c}{$cc\bar{u}\bar{d}$}&\multicolumn{2}{c}{$cc\bar{n}\bar{n}$}\\
~&$J_{3}$&$J_{4}$&$J_{1}$&$J_{2}$\\
\hline
Our work&$3.742^{+0.05}_{-0.04}$&$3.877^{+0.04}_{-0.03}$&$-$&$4.021^{+0.04}_{-0.03}$\\
Decay(GeV) &\multicolumn{2}{c}{ $D^{+}\bar{D}^{*0}(3.860\pm 0.008)$ } & \multicolumn{2}{c}{ $D^{+}\bar{D}^{*0}(3.860\pm 0.008)$ } \\
$E_{binding}$(MeV)  &$-118^{+58}_{-48}$ & $+17^{+48}_{-38}$ & $-$ & $-161^{+48}_{-38}$   \\
$[{\color{blue}6}]$&\multicolumn{2}{c}{3.978}&\multicolumn{2}{c}{4.167}\\
$[{\color{blue}34}]$&\multicolumn{2}{c}{$3.947\pm0.011$}&\multicolumn{2}{c}{$4.133\pm0.011$}\\
$[{\color{blue}5}]$&4.007&4.204&\multicolumn{2}{c}{4.201}\\
$[{\color{blue}35}]$&\multicolumn{2}{c}{4.150}&\multicolumn{2}{c}{$-$}\\
$[{\color{blue}36}]$&\multicolumn{2}{c}{4.017}&\multicolumn{2}{c}{$-$}\\
$[{\color{blue}38}]$&4.041&4.313&\multicolumn{2}{c}{4.268}\\
\hline
\end{tabular}
\end{center}

For completeness, we list in Table \uppercase\expandafter{\romannumeral4} the masses calculated with lattice QCD[{\color{blue}40}] and that in Ref. [{\color{blue}4}] for the doubly charm tetraquark $cc\bar{u}\bar{d}$ and $cc\bar{q}\bar{q}$. There, the central values correspond to $M_{b}^{2}=13\,\rm{GeV}^{2}$, ${s_{0}}(cc\bar{u}\bar{d})=19.36\,\rm{GeV}^{2}$ and ${s_{0}}(cc\bar{n}\bar{n})=20.25\,\rm{GeV}^{2}$, and the first and second uncertainties are due to the Borel parameter $M_{b}^{2}$ and the threshold parameter $s_{0}$, respectively, where $M_{b}^{2}$ ranges in $[10,15]\,\rm{GeV}^2$, $s_{0}(cc\bar{u}\bar{d})$ in $[18.9,19.8]\,\rm{GeV}^2$ and $s_{0}(cc\bar{n}\bar{n})$ in $[19.7,20.7]\,\rm{GeV}^2$. Here, the uncertainty treatment due to the parameters is same with that for the doubly bottom tetraquark states. Remarkably, the spin-weighted mass average $3802.5$ MeV for the tetraquark $cc\bar{n}\bar{n}$ agrees well with the rude sum-rule estimate $3800$ MeV with the help of the experimental mass inputs of newly-discovered resonance $T_{cc\bar{c}\bar{c}}$ and the baryon $\Xi_{cc}^{++}$ in the introduction.

As shown by the binding energies ($E_{binding}$) in Tables \uppercase\expandafter{\romannumeral1} and \uppercase\expandafter{\romannumeral2}, all of three doubly bottom tetraquarks are stable against strong and electromagnetic decay into two bottom mesons $B\bar{B}^{*}$ or $\bar{B}^{*}\bar{B_{s}^{*}}$. In the case of doubly charmed tetraquarks in Table III a $1^{+}$ $cc$-tetraquark associted with $J_{3}$ is distinctly stable against dissociation into two charmed mesons $D\bar{D}^{*}$ and one state with associted with $J_{2}$ is unstable against strong decay. The stability of one charmed tetraquarks and the strange $bb$-tetraquarks remain to be explored due to the smallness of the binding energies compared to the uncertainty.

\subsection{Summary and remarks}
Mass estimates of the $QQ\bar{n}\bar{q}$ tetraquarks composed of two heavy quarks and two light antiquarks are quite crucial to search for them experimentally and test thereby the calculational approaches employed. If the $QQ\bar{n}\bar{q}$ tetraquarks are stable against decay into two $Q\bar{q}$ mesons one may expect they are relatively long-lived and easy to be discovered. Till now, most observed candidates fit the hidden charm form $c\bar{c}n\bar{q}$, strongly decaying to $c\bar{c}$ charmonium $+$ light mesons, except for the recently-observed tetraquark $T_{cc}$ by the LHCb \cite{LHCb2021T}. The relatively smaller mass $3875$ MeV of the LHCb-observed $T_{cc}$, slightly below the $D^{*+}D^{0}$ mass threshold, remains a puzzle in the framework of compact tetraquarks as it has masses around $3.9-4.1$ GeV.

In this work, the method of QCD sum rules is used to compute the ground-state masses of the doubly heavy systems of tetraquark states $QQ\bar{q}\bar{n}$ with $J^{P}=1^{+}$ and strangeness $S=0, -1$ via careful estimates of the Borel and threshold parameters involved. We give three mass estimates for the nonstrange DH tetraquarks with flavor content $QQ\bar{n}\bar{n}$ ($Q=c,b$) and four computed masses of the tetraquark $bb\bar{n}\bar{s}$. The computed masses of the $bb$ tetraquarks lie between $10.3-10.4$ GeV for the nonstrange states and are about $10.5$ GeV for the singly strange states. Our predicted mass $3.877_{-0.03}^{+0.04}$ GeV of the nonstrange tetraquark $cc[\bar{u}\bar{d}]$ is in consistent with the measured value $3.875 \pm 0.66 _{-0.14}^{+0.11}$ GeV of the narrow state $T_{cc}$ reported by LHCb Collaboration. By the way, the weak decay widths are given for the doubly bottom tetraquarks $bb\bar{n}\bar{q}$ and compared with other calculations cited.

Our mass predictions are in agreement with the other calculations for the doubly bottom tetraquarks and slightly lower than other predictions cited for the doubly charmed tetraquarks. Combined with the weak decay widths predicted, we hope our mass predictions, with $J^{P}$ quantum numbers refined in this work, will be of helpful in searching for the doubly heavy tetraquarks or can be tested by experiments in future.

There exist some computations by QCD sum rules [{\color{blue}42,43,44,45}] of tetraquark masses, whose uncertainty hinder one to firmly claim if they are stable against strong two-meson decays. In an earlier calculation by the lattice QCD[{\color{blue}46}], the four-quark systems of doubly bottom are found to be stable, with the binding energies about $-189\pm 10\pm 3$ MeV for nonstrange systems and $-98\pm7\pm3$ MeV for the strange systems. The respective masses by recent lattice calculation[{\color{blue}47}] give the binding energies $-165\pm33 $ MeV for nonstrange systems and $-115\pm 33$ MeV for strange systems, which are not far away from our predictions $-(117-167)$ MeV for nonstrange states. Our computation indicates that all doubly-bottom tetraquarks with $J^{P}=1^{+}$ and a doubly-charmed tetraquarks associted with $J_{3}$ are stable against dissociation into two heavy-mesons, whereas a doubly-charmed tetraquarks associted with $J_{2}$ is srongly unstable. The stability of other charmed tetraquarks as well as the strange $bb$-tetraquarks remain to be undetermined.

One of main limitations in our mass computation of the $T_{cc}$ may come from ignoring the mixing effects of two-meson molecule components. In the very large limit of the $M_{Q}$, the heavy quark pair $QQ$ in tetraquark $T_{QQ}$ in $\bar{3}_{c}$ stays close to each other to form a compact core due to the strong Coulomb interaction, with the light quarks moving around the $QQ$-core \cite{Manohar:B93,MPR:D19}. In this limit, the DH tetraquark mimics a helium-like QCD-atom, for which our method in this work applies. In the finite heavy-quark limit(e.g., in charm sector), however, the $cc$ tetraquark tends to resemble the hydrogen molecule, with the scalar antidiquark $\bar{q}\bar{q}$ playing a role similar to electron (spin-singlet) pair in hydrogen molecule\cite{MPR:D19,MPR:D21}. This hints that the $T^{+}_{cc}$ may not be pure compact exotic hadron, but that of mixing state containing the molecule($DD^{*}$ or $D^{*}D^{*}$) components. The physical effects of DH tetraquark mixing of the molecule components remains to be explored\cite{Dengzhu:22}.
\section*{ACKNOWLEDGEMENTS}

D. J. is supported by the National Natural Science Foundation of China under the No. 12165017. D. G. thanks Jin-Bo Zhao for hospitality of his visiting Institute of Modern Physics, CAS. Y-J.S. is supported in part by the National Natural Science Foundation of China under the Grant No. 11365018 and No. 11375240.
\section*{APPENDIX A: Spin-color contents and the currents}
The detailed correspondence between quantum numbers and interpolating currents goes beyond our topics of this work. We confine ourself to natively discuss the spin-color contents associated with the currents $J_{1\sim4}$ to show why some configurations(denoted by $-$ in Tables I, II, and IV) do not show up.

\begin{center}
Table V: Currents and associated quantum number $J^{P}$ and the possible \\ flavor-color structures of the DH tetraquarks.
\begin{tabular}{p{3cm}<{\centering}p{5cm}<{\centering}p{3cm}<{\centering}}
\hline
$q\Gamma q$&$J^{P}$&(Flavor, Color)\\
\hline
$q_{a}^{T}C\gamma_{5}q_{b}$&$0^{+}$&($6_{f},6_{c}$),($\bar{3}_{f},\bar{3}_{c}$)\\
$q_{a}^{T}Cq_{b}$&$0^{-}$&($6_{f},6_{c}$),($\bar{3}_{f},\bar{3}_{c}$)\\
$q_{a}^{T}C\gamma_{\mu}\gamma_{5}q_{b}$&$1^{-}$&($6_{f},6_{c}$),($\bar{3}_{f},\bar{3}_{c}$)\\
$q_{a}^{T}C\gamma_{\mu}q_{b}$&$1^{+}$&($6_{f},\bar{3}_{c}$),($\bar{3}_{f},6_{c}$)\\
\multirow{2}{*}{$q_{a}^{T}C\sigma_{\mu\nu}q_{b}$}&$1^{-}$,for $\mu,\nu=1,2,3$&\multirow{2}{*}{($6_{f},\bar{3}_{c}$),($\bar{3}_{f},6_{c}$)}\\
~&$1^{+}$,for $\mu=0,\nu=1,2,3$&~\\
\multirow{2}{*}{$q_{a}^{T}C\sigma_{\mu\nu}\gamma_{5}q_{b}$}&$1^{+}$,for $\mu,\nu=1,2,3$&\multirow{2}{*}{($6_{f},\bar{3}_{c}$),($\bar{3}_{f},6_{c}$)}\\
~&$1^{-}$,for $\mu=0,\nu=1,2,3$&~\\
\hline
\end{tabular}
\end{center}

For the tetraquark system $QQ\bar{n}\bar{q}$ ($n$= $u$ and $d$, $q$ = $u$, $d$ and $s$), the pair $\bar{n}\bar{q}$ of light quarks can be either in the symmetric 6 representation(rep.) or in antisymmetric $\bar{3}$ rep. in both of the flavor and the color space. Listing all properties of diquark operators, one has correspondences between them, as in Table V. Given Table V, one may write the currents ($J_{1\sim4}$) with associated color structure of the subsystem pairs ($QQ$, $\bar{q}\bar{q}$) and their color-spin classifications(some current-color combination do not respond to any color-spin structure, denoted by "\"), as listed in Table VI and VII with $Q=b$. The situation of the doubly charmed(nonstrange) tetraquarks $cc\bar{n}\bar{n}$ is same with that of the $bb\bar{n}\bar{n}$ in Table VI.

The mass difference between that associated with $J_{1}$ and $J_{2}$, or $J_{3}$ and $J_{4}$, rise from flavour rep. and/or color rep.. As for that between the currents $J_{3}$ and $J_{4}$, other explanation may be there, e.g., the difference in form-factor of the diquarks, which is due to the different interactions within a diquark (antidiquark).

\medskip

\begin{center}
Table VI: Currents with associated quantum number $J^{P}=1^{+}$ and the possible \\ flavor-color structures of the doubly bottom(nonstrange) tetraquarks given.
\begin{tabular}{p{2cm}<{\centering}p{2cm}<{\centering}p{2cm}<{\centering}p{2cm}<{\centering}p{2cm}<{\centering}}
\hline
~&$bb\bar{q}\bar{q}$&$bb\bar{q}\bar{q}$&$bb\bar{u}\bar{d}$&$bb\bar{u}\bar{d}$\\
\hline
$J^{P}=1^{+}$&$J_{1}^{6_{c}\bigotimes\bar{6}_{c}}$&$J_{2}^{\bar{3}_{c}\bigotimes3_{c}}$&$J_{3}^{\bar{3}_{c}\bigotimes3_{c}}$&$J_{4}^{\bar{3}_{c}\bigotimes3_{c}}$\\
\hline
~&$\backslash$&$\{bb\}^{\bar{3}}_{1}\{\bar{q}\bar{q}\}^{3}_{1}$&$\{bb\}^{\bar{3}}_{1}[\bar{u}\bar{d}]^{3}_{0}$&$\{bb\}^{\bar{3}}_{1}[\bar{u}\bar{d}]^3_{0}$\\
\hline
\end{tabular}
\end{center}

\begin{center}
Table VII: Currents with associated quantum number $J^{P}=1^{+}$ and the possible \\ flavor-color structures of the doubly bottom(strange) tetraquarks given.
\begin{tabular}{p{2cm}<{\centering}p{2cm}<{\centering}p{2cm}<{\centering}p{2cm}<{\centering}p{2cm}<{\centering}}
\hline
~&$bb\bar{q}\bar{s}$&$bb\bar{q}\bar{s}$&$bb\bar{q}\bar{s}$&$bb\bar{q}\bar{s}$\\
\hline
$J^{P}=1^{+}$&$J_{1}^{6_{c}\bigotimes\bar{6}_{c}}$&$J_{2}^{\bar{3}_{c}\bigotimes3_{c}}$&$J_{3}^{\bar{3}_{c}\bigotimes3_{c}}$&$J_{4}^{\bar{3}_{c}\bigotimes3_{c}}$\\
\hline
~&$\{bb\}^{6}_{0}\{\bar{q}\bar{s}\}^{\bar{6}}_{1}$&$\{bb\}^{\bar{3}}_{1}\{\bar{q}\bar{s}\}^{3}_{1}$&$\{bb\}^{\bar{3}}_{1}[\bar{q}\bar{s}]^{3}_{0}$&$\{bb\}^{\bar{3}}_{1}[\bar{q}\bar{s}]^{3}_{0}$\\
\hline
\end{tabular}
\end{center}

\section*{APPENDIX B: The spectral densities}
The spectral densities for the current $J_{1}$ in the $QQ\bar{q}\bar{q}$ $(I=1)$ systems can be given by
$$
\begin{aligned}
\rho(s)=& \int_{x_{\min }}^{x_{\max }} d x \int_{y_{\min }}^{y_{\max }} d y \\
&\left\{\frac{(1-x-y)^{2}\left(x y s-m^{2}(x+y-4)\right)\left(m^{2}(x+y)-x y s\right)^{3}}{128 \pi^{6} x^{3} y^{3}}\right.\\
&\left.+\frac{(1-x-y)\left(m^{2}(x+y)-x y s\right)^{4}}{64 \pi^{6} x^{3} y^{3}}\right\}\\
&+3 m_{q}\langle\bar{q} q\rangle \int_{x_{\min }}^{x_{\max }} d x \int_{y_{\min }}^{y_{\max }} d y \frac{\left(m^{2}(x+y)-x y s\right)\left(m^{2}(x+y+2)-3 x y s\right)}{4 \pi^{4} x y}\\
&+\left\langle g_{s}^{2} G G\right\rangle \int_{x_{\min }}^{x_{\max }} d x \int_{y_{\min }}^{y_{\max }} d y \\
&\Bigg\{\frac{\left(m^{2}(x+y+2)-x y s\right)\left(x y s-m^{2}(x+y)\right)}{512 \pi^{6} x y}\\
&+(1-x-y)^{2}\Big[\frac{\left(m^{2}(x+y)-x y s\right)\left(m^{2}\left(3 x^{2}+3 x y-16 y^{3}+48 y\right)-5 x^{2} y s\right)}{3072 \pi^{6} x^{3} y^{2}}\\
&+\frac{m^{2}\left(2 m^{2}(x+y)+m^{2}-3 x y s\right)}{192 \pi^{6} x^{3}}\Big]\Bigg\}\\
&+\frac{m_{q}\langle\bar{q} \sigma \cdot G q\rangle\left(s-4 m^{2}\right)}{12 \pi^{4}} \sqrt{1-\frac{4 m^{2}}{s}}\\
&-\frac{2\langle\bar{q} q\rangle^{2}\left(s-4 m^{2}\right)}{9 \pi^{2}} \sqrt{1-\frac{4 m^{2}}{s}}\\
&-\frac{\langle\bar{q} q\rangle\langle\bar{q} \sigma \cdot G q\rangle}{3 \pi^{2}} \int_{0}^{1} d x\left\{\frac{m^{4}(2 x-1)}{M_{B}^{2} x^{2}(1-x)}+\frac{m^{2}(2-x)}{1-x}+M_{b}^{2} x\right\} e^{-\frac{m^{2}}{M_{b}^{2}(1-x) x}} .
\end{aligned}
$$
The spectral densities for the current $J_{2}$ in the $QQ\bar{q}\bar{q}$ $(I=1)$ systems are
$$
\begin{aligned}
\rho(s)=& \int_{x_{\min }}^{x_{\max }} d x \int_{y_{\min }}^{y_{\max }} d y\\
&(1-x-y)\left(m^{2}(x+y)-x y s\right)\left[\frac{(7-x-y)\left(m^{2}(x+y)-x y s\right)}{256 \pi^{6} x^{3} y^{3}}\right.\\
&\left.+\frac{(1-x-y)\left(m^{2}\left(3 x^{2}+6 x y-4 x+3 y^{2}-4 y-4\right)-7(x+y-1) x y s\right)}{192 \pi^{6} x^{3} y^{3}}\right]\\
&+m_{q}\langle\bar{q} q\rangle \int_{x_{\min }}^{x_{\max }} d x \int_{y_{\min }}^{y_{\max }} d y \\
& \frac{m^{2}(x+y)-x y s}{8 \pi^{4} x y}\left[2\left(m^{2}(3 x+3 y+5)-4 x y s\right)\right.\\
&\left.-(1-x-y)\left(m^{2}(-15 x-15 y+2)+25 x y s\right)\right]\\
&+\left\langle g_{s}^{2} G G\right\rangle \int_{x_{\min }}^{x_{\max }} d x \int_{y_{\min }}^{y_{\max }} d y\\
&\left\{( m ^ { 2 } ( x + y ) - x y s ) \left[\frac{(1-x-y)^{3}\left(25 x^{2} y s-m^{2}\left(15 x^{2}+15 x y-4 x+24 y\right)\right)}{9216 \pi^{6} x^{3} y^{2}}\right.\right.\\
&+\frac{(1-x-y)\left(m^{2}\left(15 x^{2} y-3 x^{2}+15 x y^{2}-7 x y+2 x-12 y\right)+x^{2}(6-25 y) y s\right)}{1536 \pi^{6} x^{3} y^{2}} \\
&\left.+\frac{(1-x-y)\left(m^{2}\left(3 x^{2}+3 x y+4 x+6 y-8\right)-x^{2}(5 x+14 y) y s\right)}{1536 \pi^{6} x^{2} y}+\frac{m^{2}(3 x+3 y+5)-4 x y s}{768 \pi^{6} x y}\right] \\
&\left.+\frac{m^{2}(x+y-1)^{2}\left(x y s(20 x+20 y-47)-m^{2}\left(15 x^{2}+x(30 y-34)+15 y^{2}-34 y+4\right.\right)}{1152 \pi^{6} x^{3}}\right\}\\
&+\frac{m_{q}\langle\bar{q} \sigma \cdot G q\rangle s}{48 \pi^{4}}\left\{\left(16 m^{2}-s\right) \sqrt{1-\frac{4 m^{2}}{s}}+\int_{x_{\min }}^{x_{\max }} d x \int_{y_{\min }}^{y_{\max }} d y \frac{6 m^{2}+2 x y s}{x}\right\} \\
&+\frac{\langle\bar{q} q\rangle^{2}\left(s+20 m^{2}\right)}{18 \pi^{2}} \sqrt{1-\frac{4 m^{2}}{s}} \\
&+\frac{\langle\bar{q} q\rangle\langle\bar{q} \sigma \cdot G q\rangle}{18 \pi^{2}} \int_{0}^{1} d x\left\{\frac{m^{4}(9-6 x)}{M_{b}^{2} x^{2}(1-x)}-\frac{2 m^{2}\left(3 x^{2}-7 x+3\right)}{x(1-x)}-\frac{3 M_{b}^{2}\left(2 x^{2}-3 x+1\right)}{1-x}\right\} e^{-\frac{m^{2}}{M_{b}^{2}(1-x) x}} .
\end{aligned}
$$
The spectral densities for the current $J_{3}$ in the $QQ\bar{q}\bar{q}$ $(I=0)$ systems are given by
$$
\begin{aligned}
\rho_{3}=& \int_{x_{\min }}^{x_{\max }} d x \int_{y_{\min }}^{y_{\max }} d y \\
& \frac{(1-x-y)\left(m^{2}(x+y)-x y s\right)^{3}\left(m^{2}((x+y)(x+y+5)-4)-x y s(x+y+1)\right)}{512 \pi^{6} x^{3} y^{3}}\\
&-3 m_{q}\langle\bar{q} q\rangle \int_{x_{\min }}^{x_{\max }} d x \int_{y_{\min }}^{y_{\max }} d y \frac{\left(m^{2}(x+y)-x y s\right)\left(m^{2}(x+y-2)-3 x y s\right)}{32 \pi^{4} x y}\\
&+\left\langle g_{s}^{2} G G\right\rangle \int_{x_{\min }}^{x_{\max }} d x \int_{y_{\min }}^{y_{\max }} d y\left\{( m ^ { 2 } ( x + y ) - x y s ) \left[-\frac{\left(x^{2}+x(8 y-2)+(y-1)^{2}\right)}{6144 \pi^{6} x^{2} y^{2}}\right.\right.\\
&\left.+\frac{6 x^{2} y\left(m^{2}(x+y-1)-2 x y s\right)-(x-y-1)\left(m^{2}\left(x^{2}+x y+4 y(y+3)\right)-2 x^{2} y s\right)}{3072 \pi^{6} x^{3}}\right] \\
&\left.+\frac{m^{2}(1-x-y)^{2}\left(m^{2}(2 x+2 y-1)-3 x y s\right)}{768 \pi^{6} x^{3}}\right\}\\
&-\frac{2m_{q}\langle\bar{q} \sigma \cdot G q\rangle\left(s+m^{2}\right)}{192 \pi^{4}} \sqrt{1-\frac{4 m^{2}}{s}} \\
&+ \frac{\langle\bar{q} q\rangle^{2} \left(s+2 m^{2}\right)}{18 \pi^{2}} \sqrt{1-\frac{4 m^{2}}{s}}\\
&+ \frac{\langle\bar{q} q\rangle\langle\bar{q} \sigma \cdot G q\rangle}{24 \pi^{2}} \int_{0}^{1} d x\left\{\frac{m^{4}}{M_{b}^{2} x^{2}(1-x)}+\frac{m^{2}(2-x)}{1-x}+M_{b}^{2} x\right\} e^{-\frac{m^{2}}{M_{b}^{2}(1-x) x}} .
\end{aligned}
$$
The spectral densities for the current $J_{4}$ in the $QQ\bar{q}\bar{q}$ $(I=0)$ systems are
$$
\begin{aligned}
\rho(s)=& \int_{x_{\min }}^{x_{\max }} d x \int_{y_{\min }}^{y_{\max }} d y\left\{\frac{(1-x-y)^{3}\left(m^{2}(x+y)-x y s\right)\left(m^{2}(3 x+3 y+1)-7 x y s\right)}{384 \pi^{6} x^{3} y^{3}}\right.\\
&\left.+\frac{(1-x-y)\left(m^{2}(x+y)-x y s\right)\left(x y s(x+y-7)-m^{2}\left(x^{2}+2 x y-3 x+y^{3}-3 y-4\right)\right)}{512 \pi^{6} x^{3} y^{3}}\right\}\\
&+ \int_{x_{\min }}^{x_{\max }} d x \int_{y_{\min }}^{y_{\max }} d y\left(m^{2}(x+y)-x y s\right)\left\{\frac{m_{q}\langle\bar{q} q\rangle\left(3 m^{2}-x y s\right)}{8 \pi^{4} x y}\right.\\
&\left.\frac{m_{q} \langle\bar{q} q\rangle\left(m^{2}\left(-15 x^{2}+x(19-30 y)-15 y^{2}+19 y+4\right)+x y s(25 x+25 y-37)\right)}{32 \pi^{4} x y}\right\}\\
&+\frac{\left\langle g_{s}^{2} G G\right\rangle}{2} \int_{x_{\min }}^{x_{\max }} d x \int_{y_{\min }}^{y_{\max }} d y\left\{\frac{m^{2}(1-x-y)^{3}\left(m^{2}(15 x+15 y+1)-20 x y s\right)}{1152 \pi^{6} x^{3}}\right.\\
&+\frac{(1-x-y)^{3}\left(m^{2}(x+y)-x y s\right)\left(25 x^{2} y s-m^{2}\left(15 x^{2}+15 x y+4 x-24 y\right)\right)}{9126 \pi^{6} x^{3} y^{2}}\\
&+\frac{(1-x-y)^{2}\left(m^{2}(x+y)^{2}-x y s\right)\left(m^{2}\left(15 x^{2} y-3 x^{2}+15 x y^{2}+x y-2 x+12 y\right)+x^{2}(6-25 y) y s\right)}{1536 \pi^{6} x^{3} y^{2}}\\
&+\frac{(x+y-1)\left(m^{2}(x+y)-x y s\right)\left(x(25 x+14) y s-m^{2}\left(15 x^{2}+x(15 y+8)+6 y+8\right)\right)}{1536 \pi^{6} x^{2} y}\\
&\left.+\frac{m^{2}(x+y-1)^{2}\left(m^{2}(6 x+6 y+1)-9 x y s\right)}{384 \pi^{6} x^{3}}+\frac{\left(m^{2}(x+y)-x y s\right)\left(m^{2}(3 x+3 y-5)-4 x y s\right)}{768 \pi^{6} x y}\right\}\\
&+ \int_{x_{\min }}^{x_{\max }} d x \int_{y_{\min }}^{y_{\max }} d y\left\{\frac{m_{q}\langle\bar{q} \sigma \cdot G q\rangle\left(3 m^{2}-x y s\right)}{96 \pi^{4} x}\right.\\
&\left.+\frac{m_{q}\langle\bar{q} \sigma \cdot G q\rangle m^{2}\left(-72 x^{2}-13 x(9 y-4)-45 y^{2}+52 y+8\right)+3 x y s(32 x+20 y-25)}{576 \pi^{4} x}\right\} \\
&-\frac{m_{q}\langle\bar{q} \sigma \cdot G q\rangle\left(104 m^{2}+2 s\right)}{1152 \pi^{4}} \sqrt{1-\frac{4 m^{2}}{s}}\\
&-\frac{\langle\bar{q} q\rangle^{2}\left(s-16 m^{2}\right)}{36 \pi^{2}} \sqrt{1-\frac{4 m^{2}}{s}} \\
&+ \frac{\langle\bar{q} q\rangle\langle\bar{q} \sigma \cdot G q\rangle}{36 \pi^{2}} \times \int_{0}^{1} d x\left\{\frac{m^{4}(12 x-9)}{M_{b}^{2} x^{2}(1-x)}-\frac{2 m^{2}(3 x-4)}{(1-x)}-\frac{3 M_{b}^{2}\left(2 x^{2}-3 x+1\right)}{1-x}\right\} e^{-\frac{m^{2}}{M_{b}^{2}(1-x) x}}
\end{aligned}
$$
The spectral densities for the current $J_{1}$ in the $QQ\bar{q}\bar{s}$ $(I=\frac{1}{2})$ systems are.
$$
\begin{aligned}
\rho(s)&=\int_{x_{\min }}^{x_{\max }} d x \int_{y_{\min }}^{y_{\max }} d y\left\{\frac{(1-x-y)^{2}\left(x y s-m^{2}(x+y-4)\right)\left(m^{2}(x+y)-x y s\right)^{3}}{256 \pi^{6} x^{3} y^{3}}\right.\\
&\left.+\frac{(1-x-y)\left(m^{2}(x+y)-x y s\right)^{4}}{128 \pi^{6} x^{3} y^{3}}\right\}\\
&+m_{q}(\langle\bar{q} q\rangle+\langle\bar{s} s\rangle) \int_{x_{\min }}^{x_{\max }} d x \int_{y_{\min }}^{y_{\max }} d y \frac{\left(m^{2}(x+y)-x y s\right)\left(m^{2}(x+y+2)-3 x y s\right)}{16 \pi^{4} x y}\\
&+\frac{\left\langle g_{s}^{2} G G\right\rangle}{2} \int_{x_{\min }}^{x_{\max }} d x \int_{y_{\min }}^{y_{\max }} d y\Bigg\{\frac{\left(m^{2}(x+y+2)-x y s\right)\left(x y s-m^{2}(x+y)\right)}{512 \pi^{6} x y}\\
&+(1-x-y)^{2}\Big[\frac{\left(m^{2}(x+y)-x y s\right)\left(m^{2}\left(3 x^{2}+3 x y-16 y^{3}+48 y\right)-5 x^{2} y s\right)}{3072 \pi^{6} x^{3} y^{2}}\\
&+\frac{m^{2}\left(2 m^{2}(x+y)+m^{2}-3 x y s\right)}{192 \pi^{6} x^{3}}\Big]\Bigg\}\\
&+\frac{m_{q}(4\langle\bar{q} \sigma \cdot G q\rangle+\langle\bar{s} \sigma \cdot G s\rangle)\left(s-4 m^{2}\right)}{192 \pi^{4}} \sqrt{1-\frac{4 m^{2}}{s}}\\
&-\frac{\langle\bar{q} q\rangle \langle\bar{s} s\rangle\left(s-4 m^{2}\right)}{9 \pi^{2}} \sqrt{1-\frac{4 m^{2}}{s}}\\
&-\frac{\langle\bar{q} q\rangle\langle\bar{s} \sigma \cdot G s\rangle+\langle\bar{s} s\rangle\langle\bar{q} \sigma \cdot G q\rangle}{12 \pi^{2}} \int_{0}^{1} d x\left\{\frac{m^{4}(2 x-1)}{M_{b}^{2} x^{2}(1-x)}+\frac{m^{2}(2-x)}{1-x}+M_{b}^{2} x\right\} e^{-\frac{m^{2}}{M_{b}^{2}(1-x) x}} .
\end{aligned}
$$
The spectral densities for the current $J_{2}$ in the $QQ\bar{q}\bar{s}$ $(I=\frac{1}{2})$ systems are
$$
\begin{aligned}
&\rho(s)=\int_{x_{\min }}^{x_{\max }} d x \int_{y_{\min }}^{y_{\max }} d y(1-x-y)\left(m^{2}(x+y)-x y s\right)\left[\frac{(7-x-y)\left(m^{2}(x+y)-x y s\right)}{512 \pi^{6} x^{3} y^{3}}\right.\\
&\left.+\frac{(1-x-y)\left(m^{2}\left(3 x^{2}+6 x y-4 x+3 y^{2}-4 y-4\right)-7(x+y-1) x y s\right)}{384 \pi^{6} x^{3} y^{3}}\right],\\
&+\int_{x_{\min }}^{x_{\max }} d x \int_{y_{\min }}^{y_{\max }} d y\left\{\frac{m_{q}\langle\bar{q} q\rangle\left(3 m^{2}+x y s\right)\left(m^{2}(x+y)-x y\right)}{8 \pi^{4} x y}\right.\\
&\left.+\frac{m_{q} \langle\bar{s} s\rangle\left(m^{2}(x+y)-x y s\right)\left(x y s(25 x+25 y-37)-m^{2}((x+y)(15(x+y)-23)+4)\right)}{32 \pi^{4} x y}\right\}\\
&+\frac{\left\langle g_{s}^{2} G G\right\rangle}{2} \int_{x_{\min }}^{x_{\max }} d x \int_{y_{\min }}^{y_{\max }} d y\\
&\left\{( m ^ { 2 } ( x + y ) - x y s ) \left[\frac{(1-x-y)^{3}\left(25 x^{2} y s-m^{2}\left(15 x^{2}+15 x y-4 x+24 y\right)\right)}{9216 \pi^{6} x^{3} y^{2}}\right.\right.\\
&+\frac{(1-x-y)\left(m^{2}\left(15 x^{2} y-3 x^{2}+15 x y^{2}-7 x y+2 x-12 y\right)+x^{2}(6-25 y) y s\right)}{1536 \pi^{6} x^{3} y^{2}}\\
&\left.+\frac{(1-x-y)\left(m^{2}\left(3 x^{2}+3 x y+4 x+6 y-8\right)-x^{2}(5 x+14 y) y s\right)}{1536 \pi^{6} x^{2} y}+\frac{m^{2}(3 x+3 y+5)-4 x y s}{768 \pi^{6} x y}\right]\\
&\left.+\frac{m^{2}(x+y-1)^{2}\left(x y s(20 x+20 y-47)-m^{2}\left(15 x^{2}+x(30 y-34)+15 y^{2}-34 y+4\right)\right)}{1152 \pi^{6} x^{3}}\right\}\\
&+\left\{\int_{x_{\min }}^{x_{\max }} d x \int_{y_{\min }}^{y_{\max }} d y \frac{m_{q}\langle\bar{q} \sigma \cdot G q\rangle\left(3 m^{2}+x y s\right)}{96 \pi^{4} x}\right.\\
&\left.-\frac{m_{q}\langle\bar{s} \sigma \cdot G s\rangle\left(m^{2}\left(72 x^{2}+117 x y+45 y^{2}-56 x-56 y+8\right)-3 x y s(32 x+20 y-25)\right)}{576 \pi^{4} x}\right\} \\
&-\frac{6 m_{q}\langle\bar{q} \sigma \cdot G q\rangle\left(20 m^{2}+s\right)+\langle\bar{s} \sigma \cdot G s\rangle\left(7 s-52 m^{2}\right)}{1152 \pi^{4}} \sqrt{1-\frac{4 m^{2}}{2}}\\
&+ \frac{\langle\bar{q} q\rangle \langle\bar{s} s\rangle\left(s+20 m^{2}\right)}{36 \pi^{2}} \sqrt{1-\frac{4 m^{2}}{s}}\\
&+ \frac{\langle\bar{q} q\rangle\langle\bar{s} \sigma \cdot G s\rangle+\langle\bar{s} s\rangle\langle\bar{q} \sigma \cdot G q\rangle}{72 \pi^{2}} \\
& \times \int_{0}^{1} d x\left\{\frac{m^{4}(9-6 x)}{M_{b}^{2} x^{2}(1-x)}-\frac{2 m^{2}\left(3 x^{2}-7 x+3\right)}{x(1-x)}-\frac{3 M_{b}^{2}\left(2 x^{2}-3 x+1\right)}{1-x}\right\} e^{-\frac{m^{2}}{M_{b}^{2}(1-x) x}} .
\end{aligned}
$$
The spectral densities for the current $J_{3}$ in the $QQ\bar{q}\bar{s}$ $(I=\frac{1}{2})$ systems.
$$
\begin{aligned}
&\rho(s)=\int_{x_{\min }}^{x_{\max }} d x \int_{y_{\min }}^{y_{\max }} d y\\
&\frac{(1-x-y)\left(m^{2}(x+y)-x y s\right)^{3}\left(m^{2}((x+y)(x+y+5)-4)-x y s(x+y+1)\right)}{512 \pi^{6} x^{3} y^{3}}\\
&+3 m_{q}(\langle\bar{s} s\rangle-2\langle\bar{q} q\rangle) \int_{x_{\min }}^{x_{\max }} d x \int_{y_{\min }}^{y_{\max }} d y \frac{\left(m^{2}(x+y)-x y s\right)\left(m^{2}(x+y-2)-3 x y s\right)}{32 \pi^{4} x y}\\
&+\left\langle g_{s}^{2} G G\right\rangle \int_{x_{\min }}^{x_{\max }} d x \int_{y_{\min }}^{y_{\max }} d y\left\{( m ^ { 2 } ( x + y ) - x y s ) \left[-\frac{\left(x^{2}+x(8 y-2)+(y-1)^{2}\right)}{6144 \pi^{6} x^{2} y^{2}}\right.\right.\\
&\left.+\frac{6 x^{2} y\left(m^{2}(x+y-1)-2 x y s\right)-(x-y-1)\left(m^{2}\left(x^{2}+x y+4 y(y+3)\right)-2 x^{2} y s\right)}{3072 \pi^{6} x^{3}}\right]\\
&\left.+\frac{m^{2}(1-x-y)^{2}\left(m^{2}(2 x+2 y-1)-3 x y s\right)}{768 \pi^{6} x^{3}}\right\}\\
&-\frac{m_{q}(\langle\bar{q} \sigma \cdot G q\rangle+\langle\bar{s} \sigma \cdot G s\rangle)\left(s+m^{2}\right)}{192 \pi^{4}} \sqrt{1-\frac{4 m^{2}}{s}}\\
&+\frac{\langle\bar{q} q\rangle \langle\bar{s} s\rangle\left(s+2 m^{2}\right)}{18 \pi^{2}} \sqrt{1-\frac{4 m^{2}}{s}}\\
&+\frac{\langle\bar{q} q\rangle\langle\bar{s} \sigma \cdot G s\rangle+\langle\bar{s} s\rangle\langle\bar{q} \sigma \cdot G q\rangle}{24 \pi^{2}}\\
&\times \int_{0}^{1} d x\left\{\frac{m^{4}}{M_{b}^{2} x^{2}(1-x)}+\frac{m^{2}(2-x)}{1-x}+M_{b}^{2} x\right\} e^{-\frac{m^{2}}{M_{b}^{2}(1-x) x}} .
\end{aligned}
$$
The spectral densities for the current $J_{4}$ in the $QQ\bar{q}\bar{s}$ $(I=\frac{1}{2})$ systems.
$$
\begin{aligned}
\rho(s)&= \int_{x_{\min }}^{x_{\max }} d x \int_{y_{\min }}^{y_{\max }} d y\left\{\frac{(1-x-y)^{3}\left(m^{2}(x+y)-x y s\right)\left(m^{2}(3 x+3 y+1)-7 x y s\right)}{384 \pi^{6} x^{3} y^{3}}\right.\\
&\left.+\frac{(1-x-y)\left(m^{2}(x+y)-x y s\right)\left(x y s(x+y-7)-m^{2}\left(x^{2}+2 x y-3 x+y^{3}-3 y-4\right)\right)}{512 \pi^{6} x^{3} y^{3}}\right\} \quad \\
&+ \int_{x_{\min }}^{x_{\max }} d x \int_{y_{\min }}^{y_{\max }} d y\left(m^{2}(x+y)-x y s\right)\left\{\frac{m_{q}\langle\bar{q} q\rangle\left(3 m^{2}-x y s\right)}{8 \pi^{4} x y}\right.\\
&\left.\frac{m_{q} \langle\bar{s} s\rangle\left(m^{2}\left(-15 x^{2}+x(19-30 y)-15 y^{2}+19 y+4\right)+x y s(25 x+25 y-37)\right)}{32 \pi^{4} x y}\right\}, \\
&+ \frac{\left\langle g_{s}^{2} G G\right\rangle}{2} \int_{x_{\min }}^{x_{\max }} d x \int_{y_{\min }}^{y_{\max }} d y\Bigg\{\frac{m^{2}(1-x-y)^{3}\left(m^{2}(15 x+15 y+1)-20 x y s\right)}{1152 \pi^{6} x^{3}}\\
&+\frac{(1-x-y)^{3}\left(m^{2}(x+y)-x y s\right)\left(25 x^{2} y s-m^{2}\left(15 x^{2}+15 x y+4 x-24 y\right)\right)}{9126 \pi^{6} x^{3} y^{2}} \\
&+\frac{(1-x-y)^{2}\left(m^{2}(x+y)^{2}-x y s\right)\left(m^{2}\left(15 x^{2} y-3 x^{2}+15 x y^{2}+x y-2 x+12 y\right)+x^{2}(6-25 y) y s\right)}{1536 \pi^{6} x^{3} y^{2}} \\
&+\frac{(x+y-1)\left(m^{2}(x+y)-x y s\right)\left(x(25 x+14) y s-m^{2}\left(15 x^{2}+x(15 y+8)+6 y+8\right)\right)}{1536 \pi^{6} x^{2} y} \\
&+\frac{m^{2}(x+y-1)^{2}\left(m^{2}(6 x+6 y+1)-9 x y s\right)}{384 \pi^{6} x^{3}}+\frac{\left(m^{2}(x+y)-x y s\right)\left(m^{2}(3 x+3 y-5)-4 x y s\right)}{768 \pi^{6} x y}\Bigg\}\\
&+ \int_{x_{\min }}^{x_{\max }} d x \int_{y_{\min }}^{y_{\max }} d y\left\{\frac{m_{q}\langle\bar{q} \sigma \cdot G q\rangle\left(3 m^{2}-x y s\right)}{96 \pi^{4} x}\right.\\
&\left.+\frac{m_{q}\langle\bar{s} \sigma \cdot G s\rangle m^{2}\left(-72 x^{2}-13 x(9 y-4)-45 y^{2}+52 y+8\right)+3 x y s(32 x+20 y-25)}{m_{q}\langle\bar{q} \sigma \cdot G q\rangle\left(96 m^{2}-6 s\right)+m_{q}\langle\bar{s} \sigma \cdot G s\rangle\left(8 m^{2}+8 s\right)}{1152 \pi^{4}}\right\} \\
&-\frac{\langle\bar{q} q\rangle \langle\bar{s} s\rangle}{\left(s-16 m^{2}\right)}{36 \pi^{2}} \sqrt{1-\frac{4 m^{2}}{s}} \\
&+ \frac{\langle\bar{q} q\rangle\langle\bar{s} \sigma \cdot G s\rangle+\langle\bar{s} s\rangle\langle\bar{q} \sigma \cdot G q\rangle}{72 \pi^{2}} \\
& \times \int_{0}^{1} d x\left\{\frac{m^{4}(12 x-9)}{M_{b}^{2} x^{2}(1-x)}-\frac{2 m^{2}(3 x-4)}{(1-x)}-\frac{3 M_{b}^{2}\left(2 x^{2}-3 x+1\right)}{1-x}\right\} e^{-\frac{m^{2}}{M_{b}^{2}(1-x) x}}
\end{aligned}
$$
with
$$
\begin{array}{ll}
x_{\max }=\frac{1+\sqrt{1-4 m^{2} / s}}{2} \quad
\\ x_{\min }=\frac{1-\sqrt{1-4 m^{2} / s}}{2} \\
y_{\max }=1-x \\
y_{\min }=\frac{x m^{2}}{x s-m^{2}} .
\end{array}
$$

\section*{APPENDIX C: Infinity of the denominator in PC}
 For the inequalities, we assume a Gedanken experiment(process): As one provides a gradually increasing energy to the tetraquark states $bb\bar{q}\bar{q}$ to produce all its excited states, some quark-antiquark pairs $Q\bar{Q}$ are created from the QCD vacuum to produce a $bb\bar{q}\bar{q}Q\bar{Q}$ hexaquark states(resonances). It is reasonable to expect that this process stops when no further higher state of the DH multiquark(hexaquark) is created via pair creation in QCD vacuum. In the case of the hexaquark produced this way, it is unknown which state of the hexaquarks $bb\bar{q}\bar{q}Q\bar{Q}$ is stable against strong decays. We assume, without loss of generality, the heaviest configuration of $bb\bar{q}\bar{q}Q\bar{Q}$ to be the $bb\bar{q}\bar{q}b\bar{b}$ in that they are stable against strong decays. For our purpose, we rest content with finding an upper limit of masses of all hexaquarks $bb\bar{q}\bar{q}b\bar{b}$ produced in this process. By (strongly) stability we assumed, there should be at least one of the $bb\bar{q}\bar{q}b\bar{b}$ hexaquark states which have the mass less than the mass sum of their final products during strong decay. Then, one infers that there are some of the hexaquark states $bb\bar{q}\bar{q}b\bar{b}$ having mass smaller than $2m_{\bar{B}/\bar{B}_{s}}+m_{\eta_{b}/h_{b}/\Upsilon/\chi_{b}}$ $= 2\cdot 5367\text{MeV}+9460\text{MeV} \approx 20\text{GeV}$. This gives the up limit of the integration in the denominator in PC.

\end{document}